\documentclass[prfluids,amsmath,amssymb]{revtex4-2}

\usepackage{physics}

\usepackage[T1]{fontenc}
\usepackage[latin1]{inputenc}
\usepackage{hyperref}
\usepackage{cleveref}
\usepackage{caption}
\usepackage{subcaption}
\usepackage{graphicx, color}

\begin{document}

\title{Experimental observations of internal wave turbulence transition in a stratified fluid.}
\author{Costanza Rodda, Cl{\'e}ment Savaro, G{\'e}raldine Davis, Jason Reneuve, Pierre Augier, Jo{\"e}l Sommeria, Thomas Valran, Samuel Viboud, and Nicolas Mordant}
\affiliation{Laboratoire des Ecoulements G{\'e}ophysiques et Industriels, Universit{\'e} Grenoble Alpes, CNRS, Grenoble-INP, F-38000 Grenoble, France}
\email{nicolas.mordant@univ-grenoble-alpes.fr}

\date{\today}

\begin{abstract}
Recent developments of the weak turbulence theory applied to internal waves exhibit a power-law solution of the kinetic energy equation close to the oceanic Garrett \& Munk spectrum, confirming weakly nonlinear wave interactions as a likely explanation of the observed oceanic spectra. However, finite-size effects can hinder wave interactions in bounded domains, and observations often differ from theoretical predictions. This article studies the dynamical regimes experimentally developing in a stratified fluid forced by internal gravity waves in a pentagonal domain. We find that by changing the shape and increasing the dimensions of the domain, finite-size effects diminish and wave turbulence is observed. In this regime, the temporal spectra decay with a slope compatible with the Garrett-Munk spectra. Different regimes appear by changing the forcing conditions, namely discrete wave turbulence, weak wave turbulence, and strongly stratified turbulence. The buoyancy Reynolds number $Re_b$ marks well the transitions between the regimes, with weak wave turbulence occurring for $1\lesssim Re_b\lesssim 3.5$ and strongly non-linear stratified turbulence for higher $Re_b$. 

\end{abstract}
\maketitle

\section{\label{sec:into}Introduction}
Internal waves, forced at near-inertial frequencies at the sea surface, are fundamental in ocean dynamics. These waves play a primary role in the energy cycle, transferring energy from tides and winds to the oceans, then propagating it towards the ocean interior, where they eventually break. It is hypothesised that mixing and dissipation occur in the deep oceans due to wave breaking \citep{wunsch2004vertical, mackinnon2017climate}. Hence, it is fundamental to fully understand internal waves to quantify the mixing and transport phenomena. 

In the open ocean, away from sources, the measured internal wave energy spectra remarkably exhibit a universal behaviour, well-described by the Garrett-Munk model, which assumes that the observed field results from a random superposition of linear internal waves \citep{garrett1979internal,polzin2011toward}. The Garrett-Munk energy spectrum, GM hereafter, relies on the assumption that the internal wave spectrum is separable in frequency $\omega$ and vertical wavenumber $m$ and gives an empirical scaling $E(m, \omega) \propto N^2 m^{-2}\omega^{-2}$ \citep{polzin2004heuristic}. Although the GM model provides a helpful description of the internal wave field, it lacks an explanation for the underlying dynamical processes and the energy transport from large to small scales. Since the 1960s, theoretical models based on weakly nonlinear wave interactions have emerged to explain the dynamics behind the ocean spectrum for surface waves \citep{hasselmann1962non}. The weak wave turbulence theory provides the mathematical framework to study wave interactions, where a kinetic equation is used to describe the energy transfer statistically through multiple waves resonances \citep{zakharov2012kolmogorov, nazarenko2011wave}. The kinetic equation has stationary solutions that predict a family of power-law decays: the Kolmogorov-Zakharov (KZ) spectra. According to this theory, energy follows a direct cascade due to the interactions between nonlinear waves. The energy flux injected at a large scale transfers towards smaller scales in analogy with the Kolmogorov energy cascade in hydrodynamic turbulence. Wave turbulence can be applied to several fields in physics, including internal waves in oceanography \citep{zakharov2012kolmogorov, nazarenko2011wave}. Internal gravity wave interactions in the weakly nonlinear regime occur among three waves, and such interactions provide a mechanism explaining the energy transfer among scales \citep{davis2020succession}. Wave turbulence theory was also applied to internal gravity waves \cite{caillol2000kinetic,medvedev2007,polzin2011toward,dematteis2021downscale}, and \ recently exhibited a candidate solution of the KZ spectrum cite{dematteis2021origins} with a scaling law $\omega^{-1.69}$, close to the empirical GM decaying slope. Other theories have been proposed to explain the dynamics leading to the observation of the GM spectrum, among which there are eikonal theories, where diffusion plays an important role \citep{Henyey_Pomphrey1983}, strongly nonlinear wave interactions, and wave diffusion by geostrophic turbulence \citep{kafiabad2019diffusion} (see \cite{staquet2002internal} for a comprehensive review).

When a stratified fluid is forced strongly, and the weak non-linearity assumption no longer holds, a strongly stratified turbulence regime can be reached \citep{lindborg2006energy, brethouwer2007}. This regime is sometimes observed near the coastal regions in the ocean, where turbulence and mixing are strong \citep{Asaro2000}. In this strongly nonlinear regime, waves play a minor role, and the energy transfers occur thanks to local instabilities of the large-scale flow, such as shear instabilities. Note that both weak and strong turbulence may occur simultaneously at different time and length scales of the flow. 

Weak wave turbulence for a random ensemble of weakly nonlinear dispersive waves is valid in the limit of unbounded domains. However, in some locations around the Earth, in laboratory experiments and numerical simulations, the propagation occurs in finite domains and shows behaviours different from those predicted by the classical wave turbulence theory. Finite-size domain effects result in a depletion of resonant triads, causing the appearance of discrete eigenmodes in the energy spectra. Intermediate states where the number of waves is significant, yet the discreteness of the wavenumber space remains important, are called discrete wave turbulence \citep{katarastova2009,l2010discrete}. The impact of dissipation and the forcing in physical systems are also often non-negligible, but they are not included explicitly in the weak wave turbulence theory. Laboratory experiments provide ideal test beds for studying wave interactions in a controlled but realistic environment. Therefore, they are of fundamental importance in investigating the limits and applicability of the weak wave turbulence theory in real systems.

Despite the abundance of theoretical studies on the analytical solutions, few laboratory experiments have been done to test the validity of theories and models in internal wave turbulence. Some examples of experimental studies on internal gravity waves leading to the observation of a turbulent cascade are \cite{brouzet2016energy, davis2020succession, savaro2020generation, Monsalve2020}. 
One of the main difficulties related to the realisation of experiments on internal wave turbulence is the size of the basin, which needs to be large enough to minimise finite-size effects and, at the same time, have non-linearities developing in the system. These conditions require the Reynolds number $Re=U\lambda_x/\nu$ to be large ($\lambda_x$ being a horizontal length scale) and, at the same time, the Froude number $F_h=U/(N\lambda_x)$ to be low. Since the viscosity, $\nu$, is fixed, and $N$ is limited in practice to order 1 rad/s values, the velocity $U$ must be relatively small. Therefore, to achieve large $Re$ for large-scale forced internal waves (typically with horizontal wavelengths comparable with the size of the domain), one needs a large experimental facility.

A first study, presented in \cite{savaro2020generation}, focuses on experiments run in the 13-meter diameter Coriolis facility in Grenoble. By forcing internal waves in a square domain, a continuum spectrum was observed, even though the eigenmodes of the box remained a signature feature of the energy spectra, indicating dominant effects due to the particular geometry of the system. 

The present study reports experiments conducted at the same Coriolis facility in Grenoble but in a pentagonal-shaped domain. We modified the domain shape to reduce the effect of discrete modes. Moreover, the pentagonal domain is larger than the square one, reducing boundary effects and increasing $Re$. Large-scale internal waves are mechanically forced in the domain, and their interactions are studied in detail to see whether they can lead to an energy cascade with characteristics similar to the one observed in the ocean and predicted by wave turbulence theory. The experiments are repeated with 20 different combinations of parameters, which are obtained by changing the amplitude and frequency of the forcing and $N$, to investigate for which buoyancy Reynolds number $Re_b$ and Froude number $Fr$ the wave turbulence regime arises. Space and time-resolved velocity measurements are used in the experiment, giving access to the details of the dynamics of the system.

The paper is organised as follows. The experimental set-up and measurement techniques are described in section \ref{sec:setup}. In section \ref{sec:domain}, we investigate the consequences of enlarging the domain and changing the shape from square to pentagon on the dynamics. Section \ref{sec:vortex} focuses on the analysis of the large-scale vortical mode that is observed to form. A detailed study of the regimes arising for different forcing frequencies is reported in section \ref{sec:forcing}. Section \ref{Dyn} discusses the regime transitions in relation to the buoyancy Reynolds number and the Froude number. Finally, we summarise our results in section \ref{sec:summary}.
%

\section{Experimental set-up}
\label{sec:setup}

The experiments presented in this study have been performed at the Coriolis facility in Grenoble. In the following, we describe the experimental set-up and the measurement techniques.

\begin{figure}

\centering
\includegraphics[width=\linewidth]{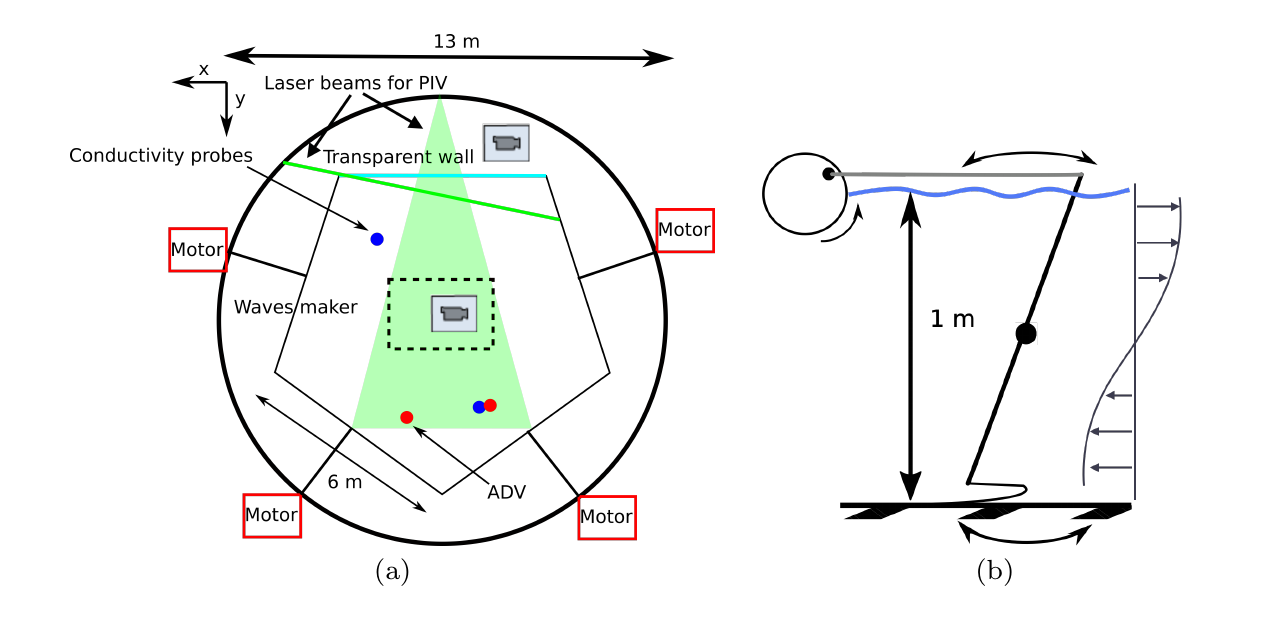}
\caption{(a) Sketch of the experimental set-up from a top view. The four walls attached to the motor are oscillating. Blue dots mark the positions of the conductivity probes and red dots that of Acoustic Doppler Velocimetry (ADV) probes. The green-coloured area also indicates the horizontal laser beam for the PIV systems. The dashed rectangle highlights the field of view of the camera. (b) Side view sketch of the moving walls used to force internal gravity waves and the resulting velocity profile along the vertical.}
\label{fig_penta}
\end{figure}

A pentagonal shape domain, consisting of five walls of six meters each, was built in a 13 meters diameter tank (see the sketch in figure~\ref{fig_penta} (a)). The tank is filled up to $H=1$ m height, with a density stratified fluid. A stable linear density profile $\rho(z)$ is created upon filling with a salt-water mixture using computer-controlled pumps. Two conductivity probes, marked by blue dots in figure~\ref{fig_penta}(a), are mounted on a vertical profiler and used to measure the density stratification through the experiments. As observed in \cite{savaro2020generation}, with a linear stratification, mixing due to the turbulent flow induced by the walls leads to the formation of homogeneous layers at the top and the bottom of the tank, while the density profile in the middle remains linear with the same slope. The tank was emptied and re-filled when the homogeneous bottom layer reached about 20 cm, which depending on the parameters of the experiments, would allow us to perform between 5 and 7 successive experiments before re-filling except for a 24 hours long experiment that was run alone (tank filled before the beginning and emptied after the end).

The linear density profile gives a constant value of the Brunt-V\"ais\"al\"a frequency, defined as 
\begin{equation}
N= \sqrt{-\frac{g}{\rho_0} \frac{d\rho}{dz}},
\end{equation}
where $\rho_0$ is the mean density. $N$ can be considered as nearly constant at the depths that are not affected by the top and mixed bottom layers, and it takes values between $0.26$ and $0.45$ rad/s depending on the experiments.\cite{savaro2020generation}.

Four vertical walls of the pentagon are attached to independent motors via crankshaft systems and oscillate around their mid-depth axes (see the sketch in figure ~\ref{fig_penta} (a), where the oscillating walls are marked by the box `motor'). The fifth wall is fixed and made of a transparent material, giving optical access to the fluid interior for measurements. Figure~\ref{fig_penta} (b) shows a sketch of one of the oscillating walls seen from the side and the vertical profile of the velocity field forced in the fluid by the oscillation of each of the walls. Both amplitude $A$ and forcing frequency $F$ can be tuned. The amplitude can be set mechanically by increments of 1~cm up to 9~cm. $F$ is defined as the dimensionless ratio between the oscillating frequency of the walls $\omega_f$ and $N$ ($F=\omega_f/N$). Amplitude and frequency are set to the same value for all moving walls, which are started simultaneously and with a random phase difference. Each oscillating wall radiates an internal gravity wave beam of finite width in the stratified fluid towards the interior of the domain, where they interact. These waves are large scale, with a vertical wavelength $\lambda_z=2$ m. The waves propagate in the interior of the fluid and are reflected by the walls, yielding a directional scatter due to the shape of the domain. In a linear regime, these reflections can lead to the buildup of resonances (linear modes) in the domain. To prevent a strong increase of the wave amplitude at resonant frequencies of the domain, the forcing frequency for each wall is slowly and randomly modulated around its average value with a maximum variation of $5\%$.

The experiments are run as follows. The walls are started simultaneously and run for 20 minutes to let the waves propagate and interact. After this time is passed, the flow is recorded for 1 hour and 20 minutes. The same procedure is repeated for all the runs presented in tables \ref{table_exps} and \ref{table_exps_all}, except `EXPlong', which was run for 24 hours.

We use the Particle Image Velocimetry (PIV) technique to measure the fluid velocity. We perform scanning three-dimensional two-component (3D2C) PIV measurements of the flow, meaning that we get the two horizontal velocity components resolved over a three-dimensional domain. The PIV set-up used is the following. 
The fluid is seeded with 200 $\mu$m  white polystyrene spheres sorted by density to match the fluid density variations with depth. The particles are illuminated by a horizontal laser sheet generated by a 25 W, 532 nm laser passing through a 60 degrees Powell lens. The depth of this laser sheet is changed using a fast oscillating mirror, which is set to scan 30 horizontal planes between $Z_{\text{min}} =0.32$ m and $Z_{\text{max}} =0.77$ m at a distance of $0.015$ m one to the other. The fluid motion is then recorded with a 12 Mpixel camera located 4 meters above the water level, which records pictures of size $2\times 3$ meters (see the dashed rectangle in figure \ref{fig_penta} (a)). The camera trigger is synchronised with the mirror's oscillation, and for each plane, the camera acquires two consecutive pictures with a 0.3 s time interval. These pairs are then used to obtain the PIV fields by calculating the seeding particle displacement between the two pictures. The interval between each pair of vertical scans is 2 seconds, defining the sampling rate. Such a period has been chosen to be small enough to resolve the fastest scales of the flow we are interested in studying. Given $N=0.45$ rad/s the fastest waves have a period $T=2 \pi/N = 14$s, hence they are well-resolved. Furthermore, we need to measure a long enough set-up time to resolve the slow scales of the flow as well (the data acquisition is limited by the capacity of the SSD drives where the data are stored). Therefore, the planes are scanned on time scales much faster than the flow, allowing time-resolved vertical measurements. After the data acquisition, the velocity field is calculated by processing the images with the open-source FluidImage tool \citep{Mohanan2018FluidSimMO} (downloadable at https://fluidimage.readthedocs.io/en/latest/). We finally obtain the two horizontal velocity components ($v_x,v_y$) resolved in time and three dimensions ($x,y,z$). The size of the searching window used in the correlation computation varies according to the image quality, with a minimum of $36 \times 36$ px$^2$ and a maximum of $54 \times 54$ px$^2$ with a 50\% overlap. This gives a horizontal resolution that ranges from $2.2$ to $3.4$ cm. The minimum and maximum particle displacements are set to 0.2 and 20 pixels, respectively. Considering the displacement and the time interval between the laser pulses, the resulting range of resolved velocities is from $4 \times 10^{-2}$ cm/s to 4 cm/s. The errors, calculated as cumulative from the main sources of errors during the PIV postprocessing, are always smaller than 5\%.

Several parameters are commonly used in the literature to distinguish among different regimes in stratified turbulent flows and quantify the relative importance of stratification \citep{brethouwer2007,bartello_tobias_2013}. For example, the Reynolds and horizontal Froude numbers used in this context are usually expressed in terms of the characteristic horizontal velocity and length scale of the flow and the kinetic energy dissipation. We need to express such non-dimensional numbers as simple parameters that we can control in our experimental set-up.

We can directly control three parameters in the experiment: the Brunt-V\"ais\"al\"a frequency $N$, the forcing amplitude $A$, and the forcing frequency $F$. These parameters can be combined with the properties of internal gravity waves to express the non-dimensional numbers. The dispersion relation of internal gravity waves, $\omega^2 = N^2/(1+(k_z/k_x)^2) $ gives the forcing horizontal length scale of the wave: by considering the vertical wavelength $\lambda_z = 2H$, where $H$ is the total fluid depth, the horizontal wavelength is $\lambda_x = 2H\sqrt{1-F^2}/F$. By considering the values of $F$ and $N$ in table \ref{table_exps}, the wavelengths of the forced waves are $\lambda_z = 2$ m and $6.1<\lambda_x<28 $ m.
The forcing velocity is $U_f = FAN$, so the Reynolds $Re$, horizontal Froude $F_h$, and buoyancy Reynolds $Re_b$ numbers can be estimated as
%
\begin{equation}
Re \equiv \frac{U_f\lambda_x}{\nu}= \frac{AN2H}{\nu}\sqrt{1-F^2}, \quad F_h\equiv\frac{U_f}{N\lambda_x}=\frac{A}{2H}\frac{F^2}{\sqrt{1-F^2}},\quad Re_b\equiv ReF_h^2=\frac{A^3NF^4}{2H\nu \sqrt{1-F^2}},
\label{eq:numbers}
\end{equation}

where $\nu\simeq 10^{-6}$ m$^2$s$^{-1}$ is the kinematic viscosity of water.
The values of the parameters characterizing the flow dynamics  for all experiments are given in \cref{table_exps} and \cref{table_exps_all}.

We have given two different values for the velocity, $U_f$ and $U_{rms}$. The root-mean-square velocity $U_{rms}$ is obtained from volume averaged energies of the measured data. The interpretation of $U_{rms}$ should be done cautiously since its value might be affected by a few dominant resonant modes of the system, which would increase it artificially. Nevertheless, we can see that $U_{rms}$ is of the same order of magnitude as $U_f$.

%
\begin{table}[htb]
\begin{ruledtabular}
{\begin{tabular}{lcccccccccc}
Name &$A$ &$N$& $$F$$& $Re$&$F_h$&$Re_b$&$U_{f}$&$U_{rms}$\\
 &(cm)&(rad/s)& & &&&(cm/s)&(cm/s)\\
\colrule
EXPlong & 5 & 0.45 & 0.68 & $3.3 \times 10^4$ & $1.6 \times 10^{-2}$ & 8.2 & 1.5 & - &\\
EXP0& 4 &0.44&0.67 &$2.6  \times 10^4$ & $1.2  \times 10^{-2}$ &3.8&1.2&0.7\\
\colrule
EXP1 & 9 &0.44&0.16 &$7.8 \times 10^4$ & $1.2  \times 10^{-3}$ &0.11&0.6&0.3\\
EXP2& 9 &0.44&0.26 & $7.6 \times 10^4$ & $3.2  \times 10^{-3}$ &0.76&1.0&0.8\\
EXP3 & 9 &0.44&0.30 & $7.5 \times 10^4$ & $4.1  \times 10^{-3}$ &1.3&1.2&0.7\\
EXP4 & 9 &0.44&0.38 & $7.3 \times 10^4$ & $7.3  \times 10^{-3}$ &3.6&1.5&1.2\\
\end{tabular}}
\end{ruledtabular}
\caption{\label{table_exps}Parameters for the experimental runs that are analyzed in part \ref{sec:domain} and \ref{sec:vortex}. $A$ is the forcing oscillation amplitude, $N$ the buoyancy frequency, $F$ the forcing non-dimensional frequency, $Re$ the Reynolds number, $F_h$ the horizontal Froude number, $Re_b$ the buoyancy Reynolds number, $U_f$ is the forced velocity, $U_{rms}$ the measured rms of the horizontal velocity (see text for definitions).}
\end{table}

\begin{table}
\begin{ruledtabular}
\centerline{\begin{tabular}{lccccccc}
Name &$A$&$N$& $F$& $Re$&$F_h$&$Re_b$ & Regime\\
 &(cm)&(rad/s)& & && & \\
\hline
N1 & 9 &0.44&0.14 & $7.8 \times 10^4$ & $8.9  \times 10^{-4}$ &0.06& DWT\\
N2 & 4 &0.45&0.41 & $3.3 \times 10^4$ & $3.7  \times 10^{-3}$ &0.45& DWT\\
N3 & 2 &0.45&0.68 & $1.3 \times 10^4$ & $6.3  \times 10^{-3}$ &0.52& DWT\\
\hline
N4 & 9 &0.36&0.29 & $6.2 \times 10^4$ &$4.0\times 10^{-3}$&1.0&WT\\
N5 & 4 &0.45& 0.54 &$3.0 \times 10^4$ & $6.9\times 10^{-3}$ &1.5&WT\\
N6 & 3 &0.45&0.68 &$2.0  \times 10^4$ & $9.5 \times 10^{-3}$ &1.8&WT\\
N7 & 9 &0.36& 0.36 & $6.0 \times 10^4$ & $6.3\times 10^{-3}$ &2.4&WT\\
N8 & 9 &0.26& 0.39 &$4.3 \times 10^4$ & $7.4\times 10^{-3}$ &2.4&WT\\
N9 & 7.5 &0.45& 0.41 &$6.1 \times 10^4$ & $6.9\times 10^{-3}$ &2.9&WT\\
\hline
N10 & 9 &0.36&0.41 & $5.9 \times 10^4$ & $8.3 \times 10^{-3}$ &4.1& ST \\
N11 & 4 &0.45& 0.68 & $2.6 \times 10^4$ & $1.3\times 10^{-2}$ &4.2& ST\\
N12 & 5 &0.26& 0.67 & $1.9 \times 10^4$ & $1.5\times 10^{-2}$ &4.4& ST\\
N13 & 5 & 0.45 & 0.68 & $3.3 \times 10^4$ & $1.6 \times 10^{-2}$ & 8.2 & ST\\
N14 & 9 &0.26&0.67 & $3.5 \times 10^4$ & $2.7 \times 10^{-2}$ &26& ST\\
N15 & 7.5 &0.45& 0.68 &$4.9 \times 10^4$ & $2.4 \times 10^{-2}$ &28& ST\\
\end{tabular}}
\end{ruledtabular}
\caption{\label{table_exps_all}Parameters for the additional experimental runs referred to in part \ref{Dyn}. $A$ is the forcing amplitude, $N$ the buoyancy frequency, $F$ the forcing frequency, $Re$ the Reynolds number, $F_h$ the horizontal Froude number, and $Re_b$ the buoyancy Reynolds number. DWT stands for discrete wave turbulence, WT for wave turbulence and ST for strongly nonlinear turbulence. {Note that experiments N4, N7, N8, N10, N12 \& N14 were conducted with mixtures of water/alcohol/salt for the purpose of optical index matching.}}
\end{table}

\section{Influence of the domain shape on the dynamics}
\label{sec:domain}

We start by summarising the previously obtained results in the square-shape domain published in \cite{savaro2020generation}. From now on, we refer to such experiments as SQExp for brevity. The SQExp investigated the flow regimes arising when internal gravity waves are forced at a fixed frequency $F = 0.7$ and amplitudes 2 to 5 cm. The study showed that increasing the forcing amplitude causes an evolution from a state of the flow dominated by the discrete modes of the square box towards a more continuous turbulent state consisting of weakly nonlinear internal waves. This first experimental evidence of weak internal wave turbulence presented some differences from what ocean observations show. The first significant difference is the dominance of discrete modes in the experiment caused by the finite size of the domain and its rectangular shape. The second difference is that the GM spectrum decaying slope $\omega^{-2}$, typically observed in oceanic spectra, could not be retrieved in the experiment. As discussed in section \ref{sec:into}, weakly nonlinear wave interaction is not the only possible cause for the observed oceanic spectra.

We propose a series of experiments introducing two main modifications to investigate whether weakly nonlinear wave turbulence can lead to energy spectra with characteristics similar to the GM spectrum. The first is changing the domain shape from a square to a pentagon. Both shapes are built with $6$ m sides; therefore the (horizontal) surface area increases from $A_s=36$ m$^{2}$ (square) to $A_p \approx 60$ m$^{2}$ (pentagon). Furthermore, in the pentagon, the number of oscillating walls, i.e. wavemakers, has been increased from 2 to 4, allowing for a roughly equivalent energy input per cubic meter of water in the domain. The second modification consists in increasing the forcing amplitude to 9 cm and lowering the forcing frequency $F$ to get closer to the forcing conditions in the ocean by tides that are very low-frequency waves (with $F\ll1$). The consequences of changing the forcing frequency are reported in section \ref{sec:forcing}.

In this section, we study how the modification of the shape and the size of the domain affects the flow. For a comparison with SQExp, we run a first experiment using the same parameters as in \cite{savaro2020generation} dataset C, i.e. forcing amplitude $A = 4$ cm, forcing frequency $F = 0.66$, 
which give $Re = 3.3 \times 10^{4}$ and $Re_b = 3.8$ (see EXP0 in table \ref{table_exps}). Note that in \cite{savaro2020generation} $\lambda_z$ is taken as $H$, so there is a factor 2 difference with the adimensional numbers calculated in this paper.
\begin{figure}
	\centering
	\includegraphics[width=0.9\linewidth]{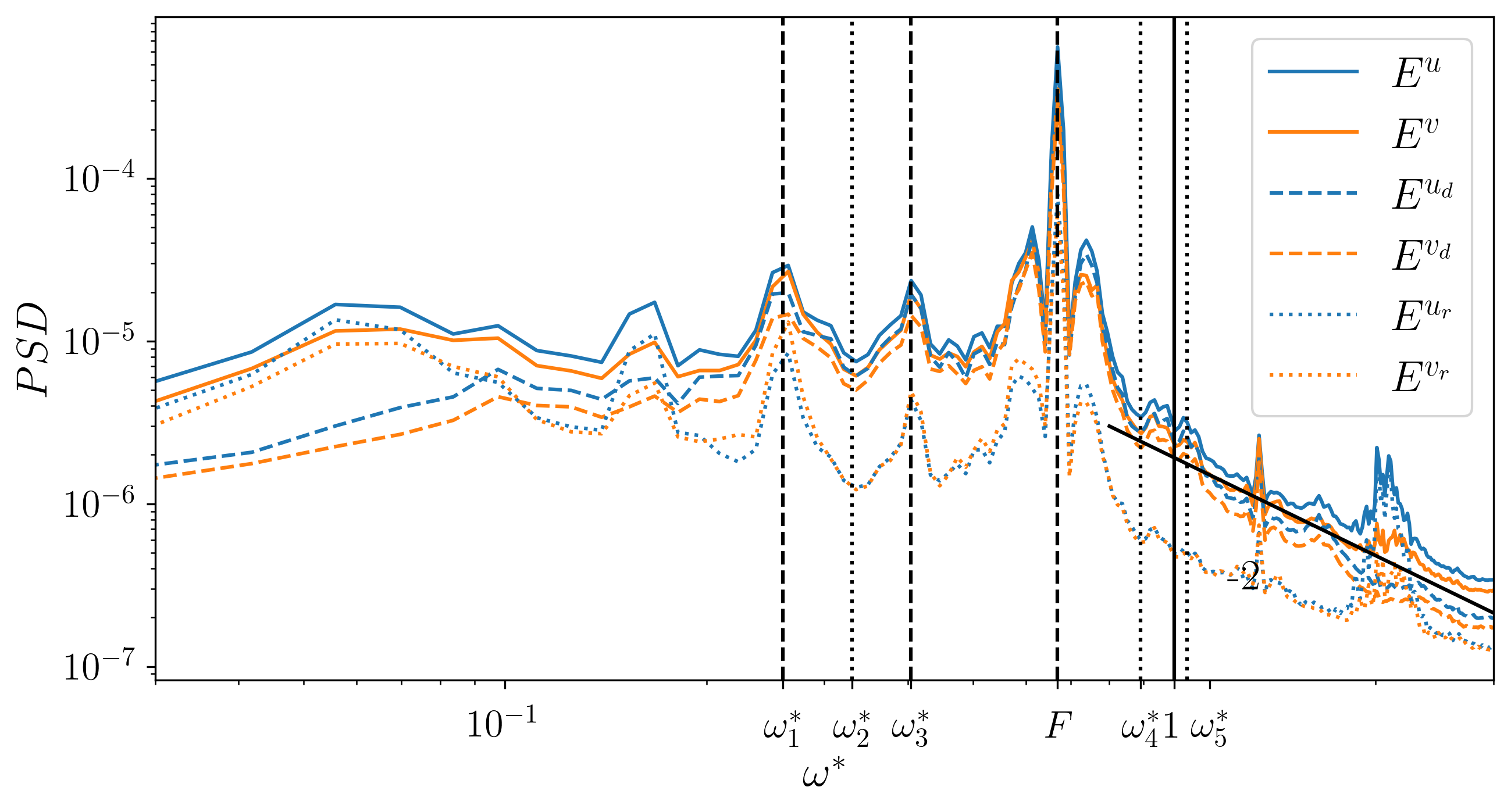}
	\caption{Power spectral density of the velocity for EXP0 as a function of the normalised frequencies $\omega^*=\omega/N$. $E^{u}$ and $E^{v}$, in solid blue and orange lines, are the spectra of the $x$ and $y$ velocity components, respectively. $E^{u_d}$ and $E^{v_d}$, dashed, and $E^{u_r}$ and $E^{v_r}$, dotted lines show the poloidal and toroidal flow decomposition. The black vertical dashed lines indicate three peaks: $\omega^*_1=0.26$, $\omega^*_3=0.40$, and the forcing frequency $F=0.67$. The black vertical dotted lines mark three off-peak frequencies: $\omega^*_2=0.33<F$, $F<\omega^*_4=0.89<1$, and $\omega^*_5=1.05>1$. These lines mark frequencies that are used for the spatio-temporal analysis in figures \ref{fig_control_autocorel_peaks} and \ref{fig_control_autocorel_waves}. The solid vertical line marks $N$. A $\omega^{-2}$ line is added to show the GM spectrum.}
	\label{fig_control_spectrum}
\end{figure}
%

\subsection{Frequency analysis}
\label{sub:freq-ref}
We begin with a temporal analysis of the velocity field developing in the experiment. Figure \ref{fig_control_spectrum} shows the frequency spectra of the horizontal velocities, averaged over the 30 vertical levels and the spatial ($x,y$) domain (see figure \ref{fig_penta} (a) for how the cartesian coordinates are defined), as a function of the normalised frequency $\omega^*=\omega/N$, $E^u$ (blue) for the $u$ component and $E^v$ (orange) for the $v$ component. The frequency spectra are calculated using a Welch method \citep{welch1967use}: the discrete Fourier transform is applied to the measured velocity time series divided into time windows of duration $T=1229$s each, with a $50\%$ overlap to improve the statistical convergence.  

To separate the internal waves from the vertical vorticity contribution to the energy spectra, we decompose the horizontal velocity field into its rotational (horizontally non-divergent) and divergent (horizontally irrotational) parts. Such decomposition is widely applied to atmospheric and oceanic flows to disentangle internal gravity waves from the rest of the flow and understand the energy contributions of each component at the various scales \citep{riley1981direct, lindborg2007stratified}. 

We use a Helmholtz decomposition to separate the horizontal velocity field $\vb{u}$ into the sum of a horizontally divergent component $\vb{u}_{\text{d}}$ and a component carrying the vertical vorticity $\vb{u}_{\text{r}}$
\begin{equation}
\vb{u}= \vb{u}_{\text{d}} + \vb{u}_{\text{r}}.
\end{equation}

The decomposition is done in practice by calculating 
\begin{equation}
\vb{u}_{\text{d}}= \mathcal F^{-1} \left[\vb{k} \left ( \frac{\hat{u}k_x+\hat{v}k_y}{k_x^2+k_y^2} \right ) \right],
\end{equation}
where $\hat{u}$ and $\hat{v}$ denote the Fourier transform of the velocities and $\mathcal F^{-1}$ the inverse Fourier transform operation. $\vb{u}_{\text{r}}$ is then simply calculated by subtracting the divergent component from the horizontal velocity field. This field separation can be interpreted as a wave-vortex decomposition, also known as poloidal-toroidal decomposition \citep{cambon2001turbulence}. In this physical interpretation, the toroidal component of the flow corresponds to vertical vortices since they are horizontally divergent-free. On the other hand, the poloidal component contains waves but also more in general unbalanced motion with horizontal vorticity, including stratified turbulence. Other techniques have been developed to obtain a wave-vortex separation for one-dimensional spectra (for example, the decomposition by \cite{buhler2014wave} that has been applied to laboratory data in \cite{rodda2020transition}). However, the poloidal-toroidal decomposition is sufficient for this study and the discussion. The divergent and rotational components are shown in figure \ref{fig_control_spectrum} in dashed and dotted lines, respectively. The poloidal components $E^{u_d}$ and $E^{v_d}$ account for most of the energy at high frequencies ($\omega^*>0.1$), suggesting a dominance of waves at most frequencies. At $\omega^*<0.1$, the toroidal components $E^{u_r}$ and $E^{v_r}$ dominate the energy, suggesting the formation of a vortical structure flow at very low frequencies. We shall investigate more in detail the large-scale dynamics in section \ref{sec:vortex}.\\

Another peculiar feature we can notice in the frequency spectra is the continuous spectrum decaying in energy for frequencies higher than the forcing. Although the forcing frequency is relatively close to $N$, the spectra in figure \ref{fig_control_spectrum} are decaying as $\approx \omega^{-2}$ for frequencies $\omega^*>0.9$. The $-2$ power law is the well-known Garrett-Munk (GM) spectrum, observed in the kinetic energy oceanic spectra \citep{garrett1979internal}. Differently from what is predicted by the GM spectra, we can observe in figure \ref{fig_control_bicoherence} that the $-2$ slope extends beyond $N$. Since internal gravity waves cannot have frequencies higher than $N$, one would expect a drop in the energy spectrum, in contrast with what is observed here. It follows that the spectrum at frequencies greater than $N$ cannot be due to propagating weakly non-linear internal waves. This super buoyancy extension is also observed in some ocean observations, as reported in \cite{polzin2011toward}.

For frequencies lower than the forcing, the spectrum is mostly flat, with only a few peaks. By comparing figure \ref{fig_control_spectrum} with the frequency spectra previously observed in figure 4a in \cite{savaro2020generation}, we can notice that the peaks at frequencies lower than $F$ are much less energetic and fewer in number. In SQExp, most of the low-frequency peaks were attributed to the 2D modes of the box with a vertical wavelength equal to $2H$ and $H$. Modes for a pentagonal domain do exist, but while for simpler geometries such as squares or rectangles, an analytical solution can be found \citep{Nurjianyan2013}, the pentagonal geometry is much more complex, and the frequencies of the modes are harder to predict.

In uniform stratification, internal waves interact through triads of Fourier modes \citep{staquet2002internal}, which satisfy the following spatial and temporal conditions
\begin{align}
\bold{k_1}+\bold{k_2}&=\bold{k_3}, \label{eq:3k}\\
\omega_1+\omega_2& =\omega_3. \label{eq:3om}
\end{align}

The most energetic frequency peak in figure \ref{fig_control_spectrum} corresponds to $F$ (third dashed vertical line counting from the left). Some other peaks appear at lower frequencies, the second and third most energetic ones at $\omega^*_1=0.26$ and $\omega^*_2=0.40$, respectively (other dashed lines in the figure). Considering the relationship between these two frequencies and the forcing frequency, we find that they satisfy the temporal resonance condition $\omega^*_1+\omega^*_3=F$, suggesting a triadic resonant interaction with the primary wave of frequency $F$ forcing two subharmonic waves, most likely resulting from a subharmonic instability \citep{joubaud2012}. Such triadic resonances have been frequently observed experimentally for inertial waves \citep{bordes2012experimental,Monsalve2020, mora2021three} and internal waves \citep{brouzet2016energy,davis2020succession}.

\subsection{Space-frequency analysis}

We complement the temporal analysis presented so far with a study in the spatial space to investigate wave interactions and dynamics. Previous studies on wave turbulence have used spatial correlation functions for this type of analysis (see \cite{campagne2015disentangling} and \cite{savaro2020generation}). The vertical extension of our measurement is limited to about 50~cm, which strongly restricts the vertical spectral resolution if one performs a Fourier analysis. The resolution is further reduced by the fact that a Hanning (or similar) window must be used. Using spatial correlation prevents this additional reduction.

The spatial correlation function of $u,v$ velocity components at fixed frequency $\omega$ is
\begin{equation}
C^u(\mathbf{r},\omega)  = \frac{\langle u(\mathbf{R_0}+\mathbf{r},\omega)u^*(\mathbf{R_0},\omega) + \text{c.c.} \rangle_{\mathbf{R_0}}}{2\langle |u(\mathbf{R_0},\omega)|^2 \rangle_{\mathbf{R_0}}},
\label{eq:autocorel}
\end{equation}
where $u$ is the temporal Fourier transform of the velocity field, $\langle \cdot \rangle_{\mathbf{R_0}}$ indicates a spatial average in $\mathbf{R_0}$, and c.c. is short for the complex conjugates. The Fourier transform in time, and the temporal average are evaluated using Welch's method to improve the statistical convergence. $C^u(\mathbf{r},\omega)$ measures the self-similarity of the signal relative to a spatial displacement for each frequency. If we now consider the correlation in the case of internal gravity waves, the flow is invariant along the phase lines. Then the autocorrelation should highlight the direction of phase lines fixed by the frequency via the dispersion relation of internal gravity waves
\begin{equation}
\omega^2 = N^2\sin^2\theta,
\label{eq:dispersion}
\end{equation}
where $\theta$ is the angle between the phase line and the horizontal axis. Similarly to what was previously done in \cite{savaro2020generation}, we consider a simple model consisting of an axisymmetric superposition of random internal gravity waves in a continuous range of wave numbers. The correlation data for each velocity component $u$ and $v$ evaluated at $ y=0$, obtained by azimuthal integration of waves numbers, is:
\begin{subequations}
\begin{align}
C^{u}(x,z,\omega) & = 2 \Bigg \langle \left[J_0(kx\sin\theta) - \frac{J_1(kx\sin\theta) }{kx \sin\theta} \right] \cos(kz\cos\theta) \Bigg \rangle_k^W, \label{eq:autocorel_model}\\
C^{v}(x,z,\omega) & = 2 \bigg \langle  \frac{J_1(kx\sin\theta) }{kx \sin\theta} \cos(kz\cos\theta) \bigg \rangle_k^W, \label{eq:autocorel_model-v}
\end{align}
\label{eq:autocorel_model-tot}
\end{subequations}
where $J_0$ and $J_1$ are the Bessel function of the first kind and $k$ is the wavenumber. $\theta$ is imposed by $\omega$ through the dispersion relation (\ref{eq:dispersion}). We used the weighted average notation
\begin{equation}
\left\langle x\right\rangle_k^W \equiv \frac{\int x(k)W(k)dk}{\int W(k)dk},
\end{equation}
with weight
\begin{equation}
W(k)\equiv k\langle |a(k,\omega)|^2\rangle,
\end{equation}
where $a(k,\omega)$ is the wave amplitude, for which we have to choose a wave spectrum model such that $\langle |a(k,\omega)|^2\rangle \propto 1/k^2$ (the choice of the exponent of the spectrum has a weak impact (see \cite{savaro2020generation}). The $\theta$ dependency in (\ref{eq:autocorel_model-tot}) is determined by $\omega$ through the dispersion relation (\ref{eq:dispersion}).

\begin{figure}
		\centering 	
		\includegraphics[width=0.9\linewidth]{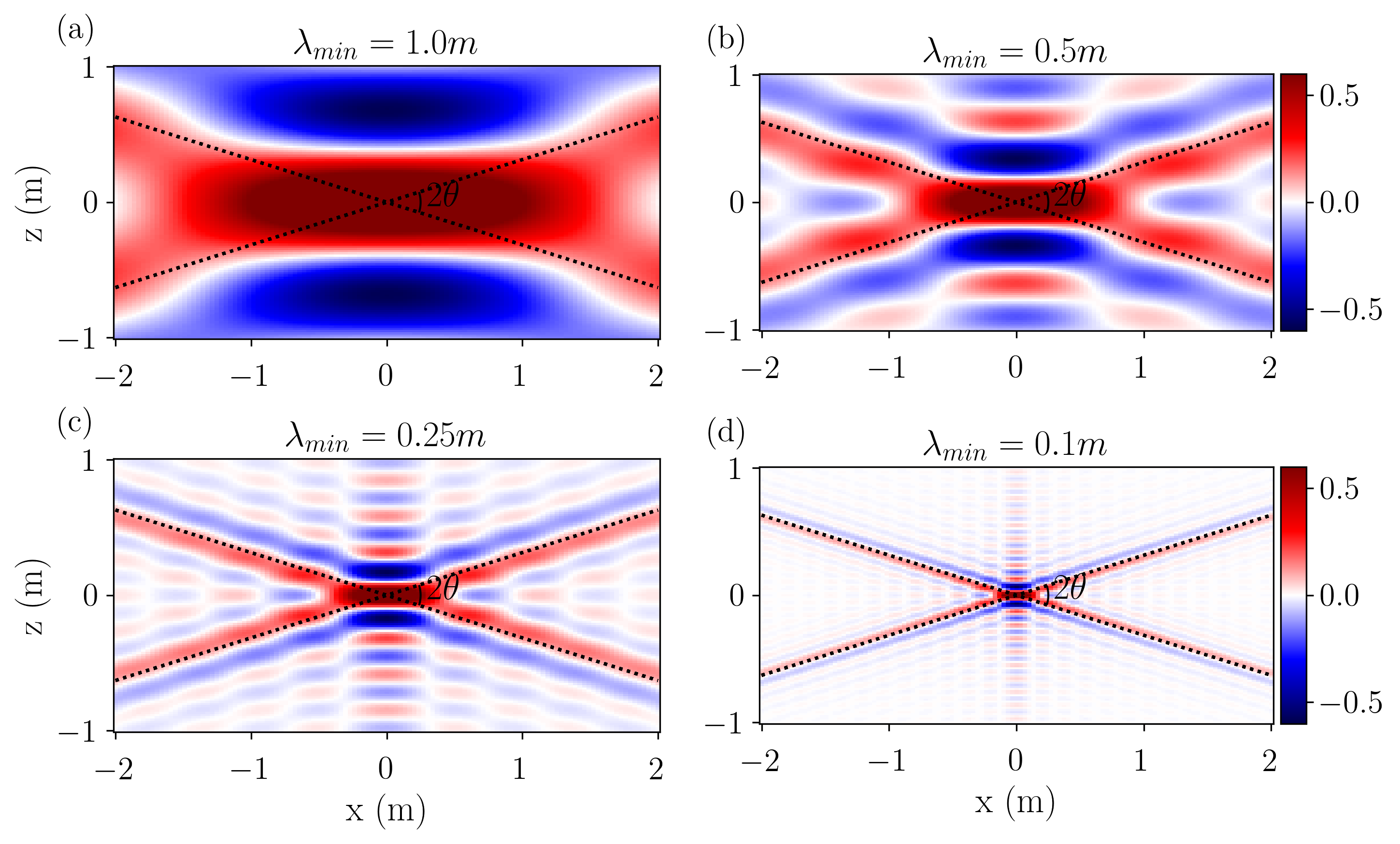}
		\caption{Numerical integration of the axisymmetric model \cref{eq:autocorel_model} for $\omega^\star=0.3$, with a continuous range of wavenumbers from $k=\pi$  up to: (a) $2\pi$, (b) $4\pi$, (c) $8\pi$, (d) $20\pi$. The decay of the $k$ spectrum is taken as $\langle |a(k,\omega)|^2\rangle \propto 1/k^2$. The black dotted line shows the internal gravity wave dispersion relation (\ref{eq:dispersion}).}
		\label{fig_model}
\end{figure}
Figure \ref{fig_model} exemplifies the effect of different integration intervals of \cref{eq:autocorel_model}, for which the range of wavenumbers is set from $\pi$ to different upper boundaries: $2\pi$, $4\pi/0.5$, $8\pi$ and $20\pi$. One can clearly observe that the dispersion relation appears in the correlation when compared with the black dotted line that marks the dispersion relation (\ref{eq:dispersion}) with the typical St. Andrew cross shape. Furthermore, one can also notice that the width of arms of the cross follows the upper bound of the range of wavenumbers (modelling the inertial range) used in the computation. The width of the arms roughly corresponds to the smallest wave scale, indicated on the plots by $\lambda_{min}$. This equivalence is convenient because it provides a way to estimate the cutoff scale of the wave spectrum. We use (\ref{eq:autocorel}) to calculate the spatial correlation for the experimental data. The experimental correlations can then be compared to the model (\ref{eq:autocorel_model-tot}), shown in figure \ref{fig_model}, which is based on the assumption of axisymmetry of the waves. Even if the experimental domain is not axisymmetric, when the flow is sufficiently non-linear, one expects that the coherence length of the waves becomes smaller than the size of the domain so that resonant mode can not develop as seen in vibrating elastic plates \citep{mordant2010} and in stratified turbulence \citep{savaro2020generation}. Furthermore, for small enough scales compared to the forcing scales,  one expects that axisymmetry is restored in the spirit of the Kolmogorov theory of turbulence. This motivates the comparison of the experimental correlations to the model.

\begin{figure}
	\centering 	
	\includegraphics[width=\linewidth]{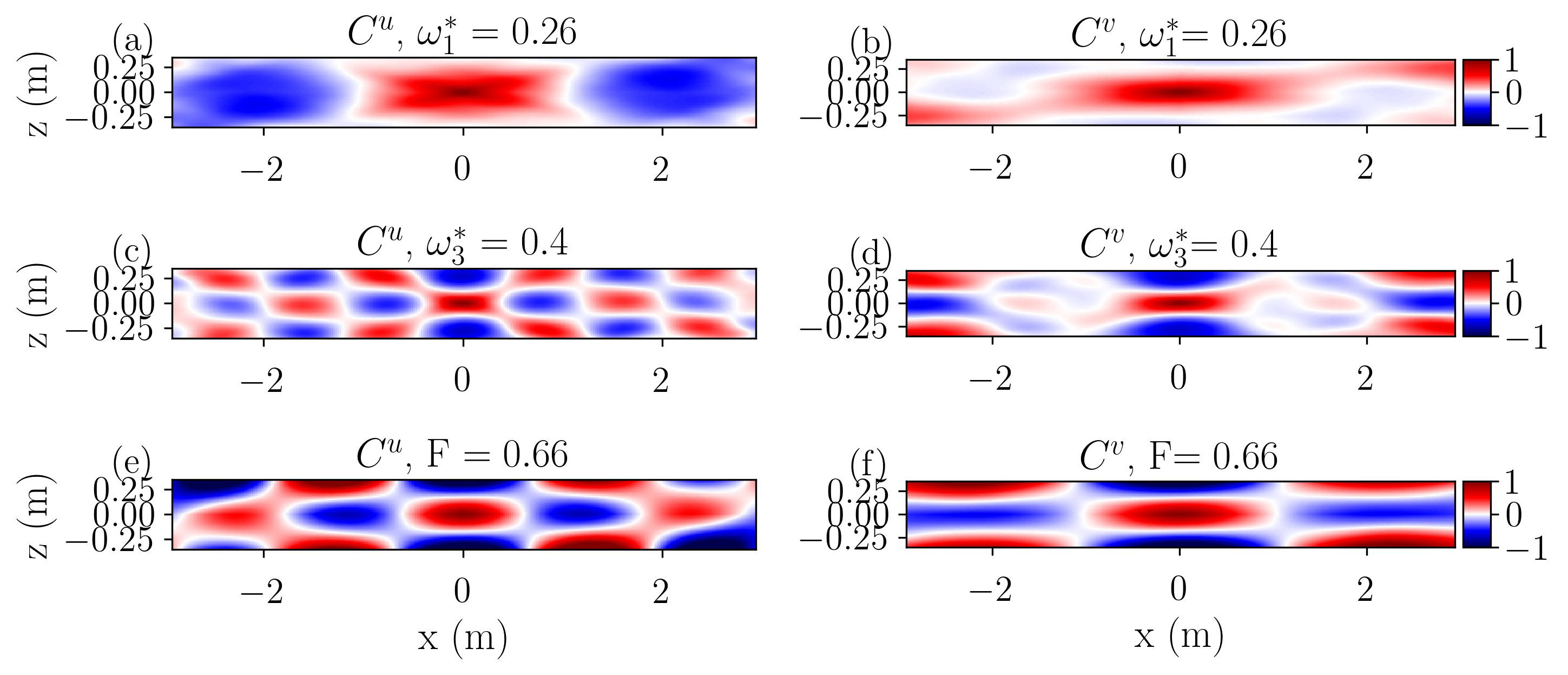}
	\caption{Space-frequency autocorrelation for EXP0 of the two velocity components $C^u$ (a,c,e) and $C^v$ (b,d,f) at three frequency peaks (marked by the vertical dashed lines in figure \ref{fig_control_spectrum}).}
	\label{fig_control_autocorel_peaks}
\end{figure}

We begin investigating the spatial structures for the three main frequency peaks highlighted in figure \ref{fig_control_spectrum} (vertical dashed lines), which we speculated might be in a triadic resonance. The correlation for these frequencies, $\omega_1^*=0.26,\, \omega_3^*=0.4,\, F=0.66$ is shown in figure \ref{fig_control_autocorel_peaks} (a), (c), and (e), for the $u$ velocity, and (b), (d), and (f) for the $v$ velocity component. Note that $u$ is the in-plane component, while $v$ is perpendicular to the vertical plane. The statistical convergence for the across-plane velocity is much slower to obtain, and therefore $C^{v}$ is less accurate than $C^{u}$.
The correlation at all the peak frequencies shares similar characteristics, with a well-recognisable chessboard pattern. These structures are typical of waves with a single dominant wavelength. \cite{savaro2020generation} identified these patterns with the 3D modes of the square domain. Here they correspond to the modes of the pentagon. From the plots of $C^u$, we can calculate the $k_x$ component of the wavevector by measuring the peak-to-peak distance. The $x$-projection of $\bold{k}$ has values $k_{x1}=3.1$ rad/m, $k_{x3}=7.8$ rad/m, and $k_{xF}=4.8$ rad/m. Combining these values, we retrieve the relation $k_{xF}=k_{x3}-k_{x1}$, compatible with the triadic resonance condition. The other two components of the wavevector are harder to measure, particularly $k_z$, because we only measure a small portion of the total fluid depth. A rough estimate for the $y$ wavenumbers gives $k_{y1}=2$ rad/m, $k_{y3}=5.2$ rad/m, and $k_{yF}=3.1$ rad/m. And for the vertical wavenumbers gives $k_{z1}<10.5$ rad/m, $k_{z3}=31.4$ rad/m, and $k_{zF}=21$ rad/m. Combined together, $k_{yF}\approx k_{y3}-k_{y1}  $ and $k_{zF}=k_{z3}-10.4$, which is not incompatible with the partial measurement of $k_{z1}$. Therefore, the spatial autocorrelation reveals wavelengths consistent with the hypothesis of triadic resonance of large-scale internal waves. The chessboard pattern can thus be interpreted as modes of the pentagon fed by triadic resonances.

\begin{figure}
	\centering 	
	\includegraphics[width=\linewidth]{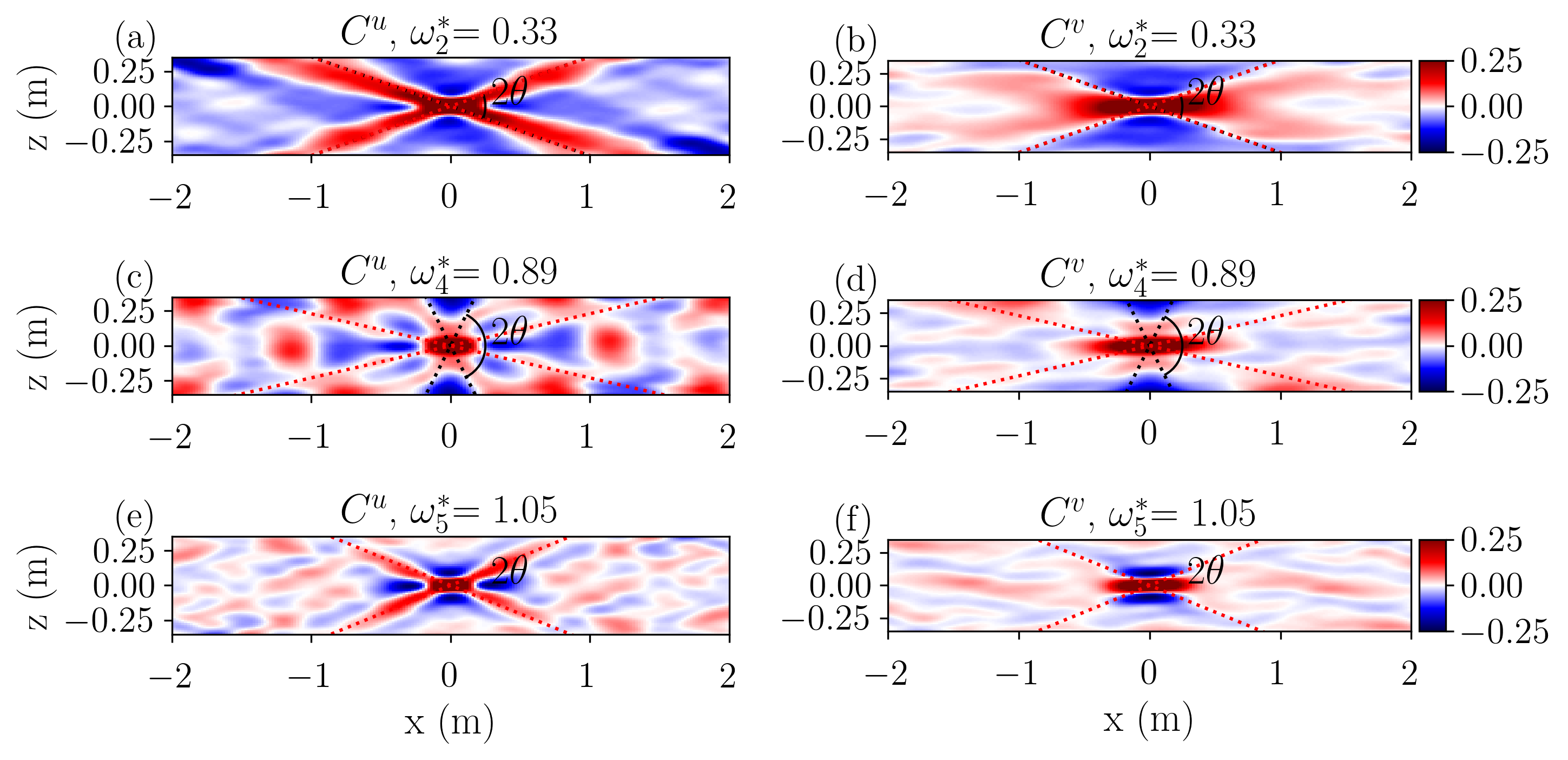}
	\caption{Autocorrelation $C^u$ (a,c,e) and $C^v$ (b,d,f) at the three off-peak frequencies in EXP0 (marked by the vertical dotted lines in figure \ref{fig_control_spectrum}). The black dotted line shows the dispersion relation (\ref{eq:dispersion}), and the red dotted line is the dispersion relation for frequency $\omega^*-F$ corresponding to bound waves (see text).}
	\label{fig_control_autocorel_waves}
\end{figure}
Off-peak frequencies are characterised by completely different patterns in the correlation, as shown in figure \ref{fig_control_autocorel_waves}. The correlation is plotted for three frequencies in the continuum of the spectra, marked by the vertical dotted line in \ref{fig_control_spectrum}. In figures \ref{fig_control_autocorel_waves} (a) and (b) the correlation $C^u$ at $\omega^*=0.26$, which is below $F$, assumes the typical the St. Andrews cross with arms at an angle $\theta$ with the $x-$axis determined by the dispersion relation \ref{eq:dispersion}. The signal for $C^v$ is weaker and less clear due to low statistical convergence. For frequencies $F<\omega^*<1$ two cross patterns can be identified, as for example shown in \ref{fig_control_autocorel_waves} (c) and (d) at $\omega^*=0.89$. One cross pattern matches the dispersion relation (black dotted line), while the second one matches the dispersion relation but for the frequency $\omega^*=0.89-F$ (highlighted by the red dotted line). {This particular feature beyonds to the category of  \textit{bound waves}. 
Very generally a quadratic linearity can transfer energy at $\omega=\omega_1\pm \omega_2$ and $\mathbf{k}=\mathbf{k_1}\pm\mathbf{k_2}$ when energy is present at $(\omega_1,\mathbf{k_1})$ and $(\omega_2,\mathbf{k_2})$ in the space-time Fourier space. If $(\omega_1,\mathbf{k_1})$ and $(\omega_2,\mathbf{k_2})$ are two waves (on the linear dispersion relation), $\omega=\omega_1\pm \omega_2$ and $\mathbf{k}=\mathbf{k_1}\pm\mathbf{k_2}$ may not be on the dispersion relation. In the field of surface waves, when it is not the case, the latter mode is called ``bound mode'' resulting for so-called ``non resonant interaction'' (see \cite{Gibson2007} for instance). We keep the same wording of ``bound wave'' in the following.} For very weak turbulence, one expects that only resonant waves are coupled so that a significant energy transfer occurs only if $(\omega,\mathbf{k})$ lies on the dispersion relation (free wave). However, for a finite level of nonlinearity, energy can also be transferred efficiently in the non-resonant case so that $(\omega,\mathbf{k})$ is not a free wave but a bound wave. This is often the case for surface gravity waves \citep{nazarenko2010statistics,herbert2010observation,campagne2019} and it has also been observed for internal waves \cite{davis2019}. The most energetic waves often correspond to $(\omega_1,\mathbf{k_1})$ being the peak of the spectrum (typically the forcing). In our case, it means $\omega_1^\star=F$ and $k_1$ correspond to a very large scale wave of length scale larger than 2~m. At scales $10-20$cm, $k_2\gg k_1$ so that $\mathbf k\approx \mathbf k_2$. Thus a bound wave occurring at a frequency $\omega^\star=F\pm\omega_2^\star$ will have the wavevector $\mathbf k\approx\mathbf k_2$ and it will exhibit the same dispersion relation as that at frequency $\omega^\star_2=\pm(F-\omega^\star)$. The correlations shown in figure ~\ref{fig_control_autocorel_waves}(c) mean that the angle of the St Andrew cross will have the angle corresponding to the reduced frequency $0.89-F$.

Interestingly, the signature of bound waves can be identified for frequencies above $N$, as figure \ref{fig_control_autocorel_waves} (e) and (f) shows. This signature is a persistent feature characterising the correlation at all frequencies above $N$ with decreasing intensity until it fades away. This suggests that the dynamics underlying the $-2$ decaying slope above $N$ are partially due to bound waves.
Finally, the arms of the cross for the lower frequencies are wider ($\approx 0.2$ m) than the ones for higher frequencies ($\approx 0.1$ m), indicating that the cutoff of the wave spectrum occurs at larger scales for lower frequencies. This suggests that the transition from waves to vortices occurs at larger scales for low frequencies where nonlinearities may be stronger. \\

\subsection{Nonlinear coupling analysis}
The frequency spectrum and the spatial correlation are insufficient to investigate the nonlinear wave interactions, as they lack information about the phase-coupling. To detect nonlinearities such as quadratic coupling among waves, we need to use higher-order spectral analysis, namely the bicoherence of the velocity field, which is defined as:
\begin{equation}
	B^u(\omega_1,\omega_2) = \frac{\langle u(\mathbf{r},\omega_1)u(\mathbf{r},\omega_2)u^+(\mathbf{r},\omega_1+\omega_2)\rangle_{\mathbf{r}}}{\left (\langle |u(\mathbf{r},\omega_1)|^2\rangle_{\mathbf{r}}\langle |u(\mathbf{r},\omega_2)|^2\rangle_{\mathbf{r}}\langle |u(\mathbf{r},\omega_1+\omega_2)|^2\rangle_{\mathbf{r}}\right)^{1/2}},
	\label{eq:bicoherence}
\end{equation}
where $^+$ denotes the complex conjugate and $\langle\rangle_{\mathbf{r}}$ indicates a spatial averaging over the measurement field.
The bicoherence measures the phase correlation of the signal between the frequencies of the two interacting waves $\omega_1$, $\omega_2$ and the resulting interaction $\omega_3=\omega_1+ \omega_2$. Random phase waves have a bicoherence close to zero. Finite values of $B^u$ highlight a non-linear coupling. For weak wave turbulence, the values of $B^u$ remain small. We detrended the velocities by subtracting their temporal average before computing the bicoherence. Furthermore, we calculated a reference level of bicoherence by adding a random phase at each spatial coordinate at the numerator in (\ref{eq:bicoherence}). Values smaller than ten times the reference are considered non-physical and therefore discarded. 
\begin{figure}
	\centering
	\includegraphics[width=0.59\linewidth]{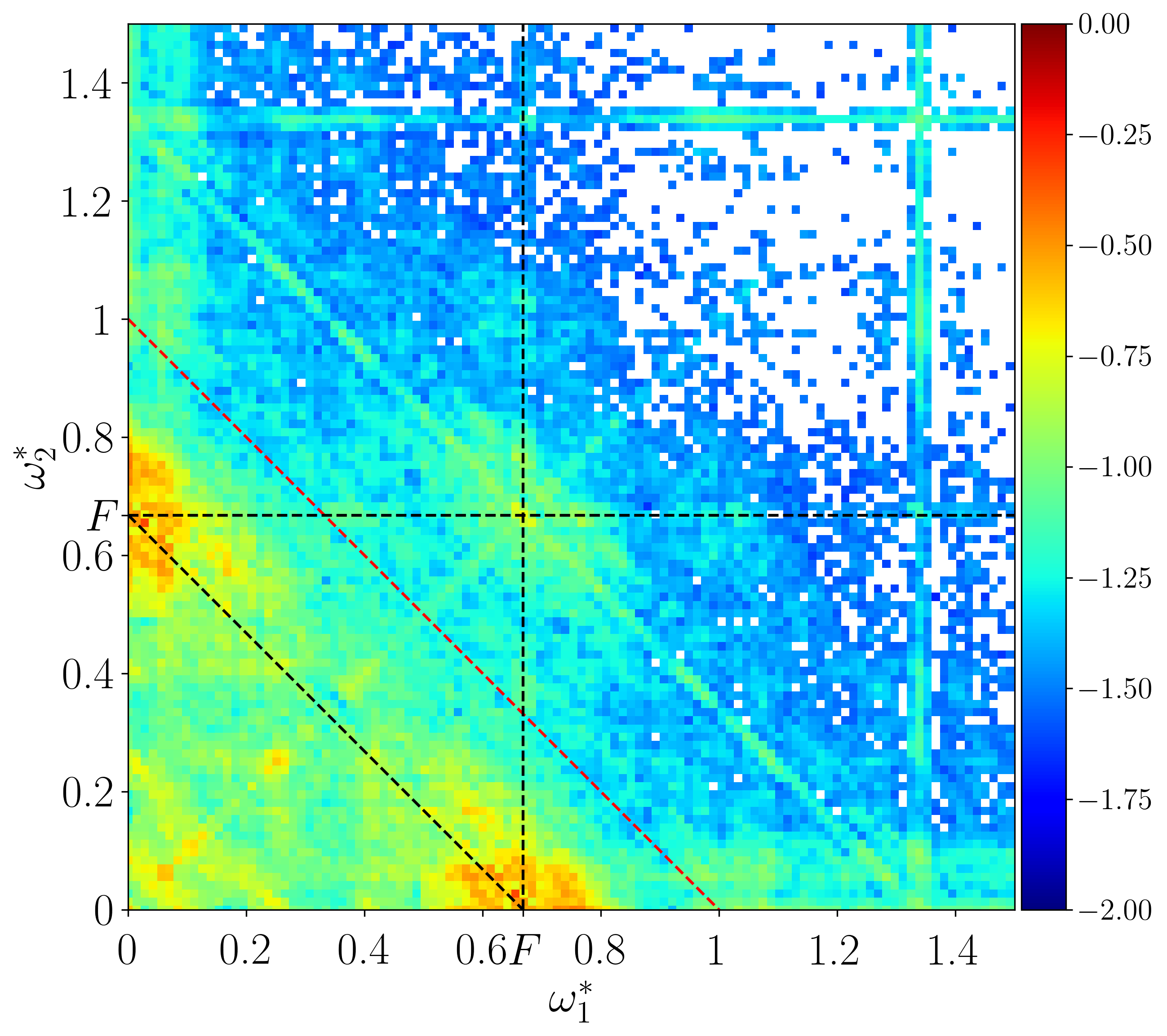}
	\caption{Color plot of the bicoherence $B^u(\omega_1,\omega_2)$ for EXP0 (colors in base-10 logarithmic scale). The black dashed line marks triadic resonances for which one of the frequencies equals $F$.  The red dashed line corresponds to $\omega_1+\omega_2=N$. Above this line, one of the three frequencies involved in the triad is not a quasi-linear wave.}
	\label{fig_control_bicoherence}
\end{figure}

Figure \ref{fig_control_bicoherence} shows the bicoherence's colour plot for EXP0. Note that the plot is symmetric by construction with respect to the bisector (the $\omega^*_1=\omega^*_2$ line). As we can see for the forcing frequency $F$, the self-correlation peaks are placed on the bisector. The possible triads related to the forcing frequency appear along the $-1$ slope highlighted by the black dashed line in figure \ref{fig_control_bicoherence}. We can see that the forcing internal wave interacts more strongly with a very low frequency ($\omega^*<0.1$) by the broad orange peaks in figure \ref{fig_control_bicoherence}. Apart from these peaks, the bicoherence for EXP0 is a smooth function, as expected for wave turbulence kinetic regimes \citep{Monsalve2020}. The peaks at $\omega^*_1=0.26$ and $\omega^*_2=0.40$ that seem significant in the frequency spectrum (in figure \ref{fig_control_spectrum}) are not particularly visible in the bicoherence, suggesting that the dynamics of EXP0 is dominated by multiple wave interactions rather than single triads. At frequencies $\omega>N$, corresponding to the upper right part in figure \ref{fig_control_bicoherence}, the only visible interaction is related to the second harmonic of the forcing, $2F$. Apart from that, we have no other prominent wave interactions for such high frequencies, and the bicoherence is dropping fast for $\omega^*>1$ (above the red dashed line)--the analysis of the bicoherence complement the spatio-temporal discussion. The global picture for this experiment is that large-scale waves interacting through triads initiate a cascade of wave interactions that lead to wave turbulence. Smaller scales and higher frequencies are dominated by a mixture of turbulence and bound waves, sustaining a $-2$ slope in the frequency spectra.

\section{Large-scale dynamics}
\label{sec:vortex}
\begin{figure}[t]
	\centering
	\includegraphics[width=\linewidth]{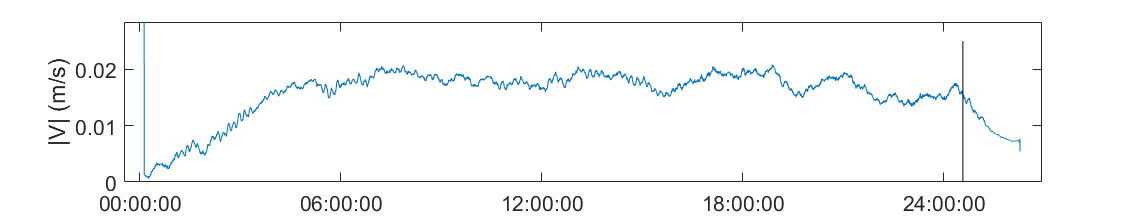}
	\caption{Temporal evolution velocity magnitude over a 24-hours long run. The data are measured with an ADV device placed at the position corresponding to the red dot on the right in figure \ref{fig_penta} at a depth $z=0.58$ m. The velocity has been smoothed over a sliding window of the duration of 900 seconds. The black vertical line marks when the wavemakers are stopped in the experiment.}
	\label{fig_vortex-vel}
\end{figure}
In section \ref{sub:freq-ref}, the spectra toroidal/poloidal decomposition gave a hint of the formation of a very slow vortical mode. 
To further investigate this vortex, a 24-hours long experiment was performed (EXPlong in table \ref{table_exps}), where the velocity is measured between $Z_{\text{min}}=0.1$ m and $Z_{\text{max}}=0.9$ m. In this run, the velocity increases linearly during the first six hours, after which it reaches a stationary state, and it remains constant for as long as the wavemakers are in action (see figure \ref{fig_vortex-vel}). 

\begin{figure}[t]
	\centering
	\includegraphics[width=0.8\linewidth]{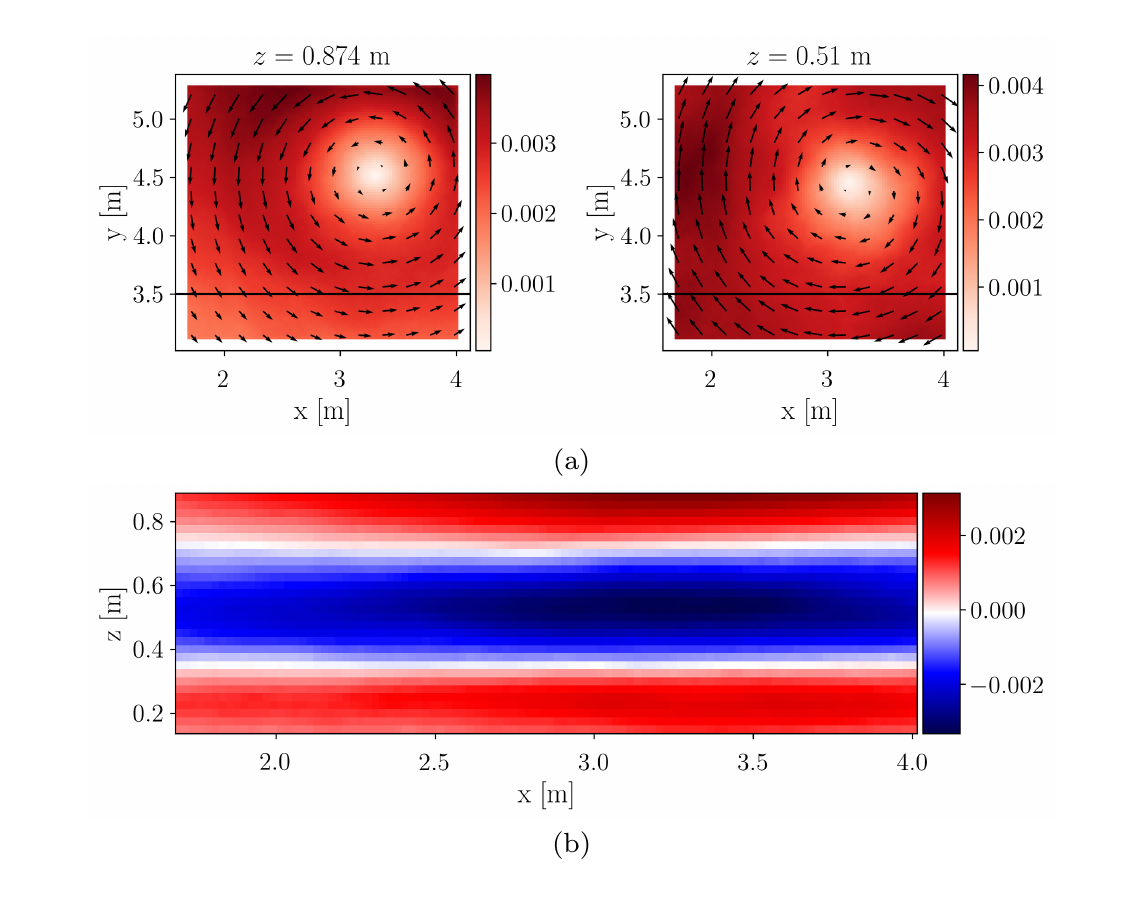}
	\caption{(a) Temporally averaged velocity field obtained by PIV in the horizontal plane at two fluid heights $z=0.874$ m (left) and $z=0.51$m (right), showing the large-scale vortex changing the direction of rotation with $z$. The colour indicates the magnitude of the velocity $U=\sqrt{u^2+v^2}$ in m/s. (b) Temporally averaged $u$ velocity component in the vertical plane $x-z$ taken at $y=3.5$ m (black horizontal line in the top plots). The red colour shows positive velocities, i.e. towards the right side of the figure, while the blue shows negative velocities, i.e. towards the left side of the figure.}
	\label{fig_meanvelocity}
\end{figure}

The time-averaged velocity field in the horizontal plane, shown in figure \ref{fig_meanvelocity} (a), shows a vortex of the size of the entire image, which covers only a small part of the whole experimental domain (see figure \ref{fig_penta}). A similar structure has also been observed in numerical simulations with geometry and forcing analogous to the lab experiments in the square geometry~\citep{reneuve2021}.

The vortex in the left plot rotates counterclockwise, while in the figure in the right, the vortex rotates clockwise. It can also be seen that the centre of the vortex shifts its position slightly with height, migrating towards the right-bottom corner of the figure for lower depths. 

The switch in the direction of rotation of the vortex is better visible in panel (b) of figure \ref{fig_meanvelocity}. The plot shows the horizontal velocity component $u$ in a vertical plane $x-z$ taken at $y=3.5$ m, marked by a solid black line in figure \ref{fig_meanvelocity}(a). The red colour belongs to positive velocities, pointing rightward in the figure, and the blue is for velocities pointing in the opposite direction. The vertical vortex structure, with inversions in the direction of rotation at two depths, is well visible. In the upper and lower parts, close to the surface and the bottom, it rotates counterclockwise; in the middle part, it rotates clockwise between $0.35$ and $0.7$ m. The vortex vertical wavelength is $\lambda_z \approx 0.6$ m.

The vortex, being a large-scale horizontal flow with a vertical structure, resembles the shear modes with zero horizontal wave vector due to the upscale energy transfer from the forcing scales observed in stratified turbulence numerical simulations \cite{maffioli2020signature,lam2021}. A mechanism generating energy transfer from the scales at which energy is injected towards the larger scales is the flux loop \citep{boffetta2011flux}. Such a mechanism consists of an inverse flux of kinetic energy and a forward flux of potential energy, which has also been observed in 2D numerical simulations of stratified turbulence forced by internal gravity waves \citep{calpelinares2020}. For rotating turbulence, a similar mechanism called bidirectional cascade is observed to lead to the formation of an energy condensation at large scales \citep{alexakis2018cascades, van2020critical}.

Note that the shear modes arising in a periodic domain, with $k_x=0$, are not physically possible in the experiment due to the presence of the lateral walls. A domain-sized vortex can arise in a bounded domain due to spectral condensation and enhance upscale energy cascade, as observed experimentally by \cite{xia2011upscale}. Because of their similar nature, we speculate that the vortex observed in our experiment is also generated by large-scale energy accumulation in an inverse energy transfer from the fast wave modes to the slow vortex mode, which is the largest mode fitting into the domain. 
Another possible explanation we cannot exclude for the vortex formation is related to the mechanical panels forcing the waves in the experiment since they inject and dissipate energy at large and intermediate scales (see \cite{reneuve2021} for more details). \\

\section{Dynamics evolution upon changing forcing frequency}
\label{sec:forcing}
This section investigates the effect of changing the wave forcing frequency on the dynamics developing in the system. We run a series of experiments with the same stratification ($N = 0.44$ rad/s), fixed forcing amplitude ($A = 9$ cm) but changing the forcing frequency from $F=0.16$ to $F=0.38$ (see details of EXP1, EXP2, EXP3, and EXP4 in table \ref{table_exps}). Because of the properties of internal gravity waves (following the dispersion relation (\ref{eq:dispersion})), the forcing frequency determines the wave angle of propagation. Waves forced at low frequency have phase speed mostly vertical (though their group speed, and energy propagation, is mostly horizontal), while waves with higher frequency have steeper phase speed. In section \ref{sec:domain} and in \cite{savaro2020generation}, waves are forced at $\theta=45^{\circ}$, which is the most effective angle for experimentally forced waves, since it maximises their vertical transport of horizontal momentum \citep{dohan2003internal}. When the forcing frequency decreases, waves are generated at a smaller angle, i.e. closer to the horizontal, which is less efficient. The reason for decreasing the forcing frequency is that in the ocean, waves are mostly forced at frequencies near inertia $f=2 \Omega$, and $f<<N$. Since we do not consider rotation in the experiments presented here, we aim to force the waves at a frequency significantly lower than $N$. Furthermore, decreasing the forcing has the advantage of enlarging the frequency range between ${F}$ and $N$, to observe a possible energy cascade to high frequencies \citep{dematteis2021downscale}.

Looking at the non-dimensional numbers in table \ref{table_exps}, it emerges that increasing the forcing frequency results in a substantial increase in the Froude and buoyancy Reynolds number (both directly proportional to the forcing $F^2$ and $F^4$ respectively). In contrast, the Reynolds number remains close to $7.10^4$.  \cite{Monsalve2020} observed a transition from a discrete-wave-interaction regime to a weak turbulence regime in the case of inertial waves caused by the increase of the Reynolds number, differently from what we observed here where the transition is obtained for increasing $Re_b$.

\subsection{Frequency analysis}
\begin{figure}[t]
	\centering
	\includegraphics[width=0.9\linewidth]{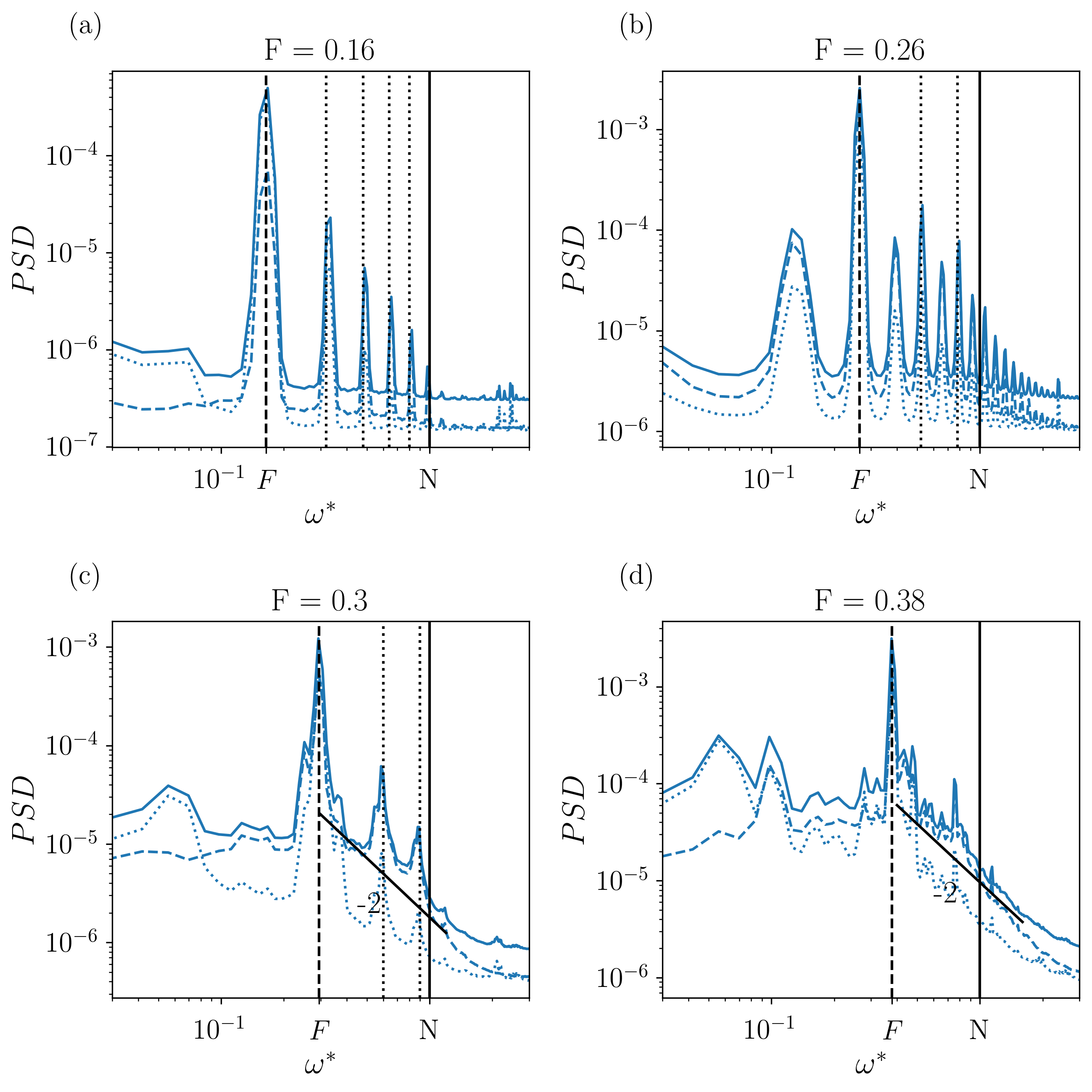}
	\caption{Power spectra density of $E=E^u+E^v$ for experiments at different forcing frequencies: (a) EXP1, (b) EXP2, (c) EXP3, (d) EXP4. The solid line indicates the total PSD, the dashed line the poloidal component, and the dotted line the toroidal component. The vertical solid line mark the frequency equal to $N$ ($\omega^*=1$), and the vertical dashed line marks the forcing frequency ($\omega^*=F$). The dotted vertical lines in (a)-(c) mark higher harmonics of the forcing ($2F$, $3F$,...).}
	\label{fig_PSD}
\end{figure}

In the following, we apply the analysis presented in section \ref{sec:domain} to EXP1 to EXP4 to study the different dynamics developing in the flow. We start the analysis with the frequency spectra in figure \ref{fig_PSD}. The most striking feature is the qualitative difference between plots \ref{fig_PSD}(a),(b) and \ref{fig_PSD}(c),(d). The first two show a discrete set of peaks typical of linear wave regimes. At the same time, the latter two have a continuous spectrum with a decay above $F$, typical of more complex wave fields where nonlinear interactions occur and are most commonly observed in the ocean. Although discrete spectra are not commonly observed in the ocean, some locations, for example, the Near Eastern boundaries of oceans \citep{van2007high}, display a discrete energy spectrum with super harmonics, similar to the ones we observe in the experiment.

A closer look at each plot can reveal other differences. For EXP1, with the smallest forcing frequency $F=0.16$, the spectrum in figure \ref{fig_PSD}(a) reveals energy peaks at the forcing frequency ($F$) and its higher harmonics $nF$, with $n=2,3$ and so on. 

The first harmonic results from the interaction of the waves forced by the walls, which are then reflected by the geometrical boundaries and interact nonlinearly \citep{tabaei2005nonlinear,Jiang2009}.
If the first harmonics has enough energy, it can interact with the forced wave and generate the third harmonic, and the process is then repeated for higher harmonics. The energy is in this way transferred from the forcing to higher frequencies without showing a continuous energy cascade since the energy level between the peaks is almost at the noise level. This energy transfer is visible in the decay of the height of the peak in the frequency spectra. \cite{husseini2020experimental} also observed the generation of superharmonics due to triadic resonances among internal waves in laboratory experiments.

\begin{figure}[t]
	\centering
	\includegraphics[width=0.7\linewidth]{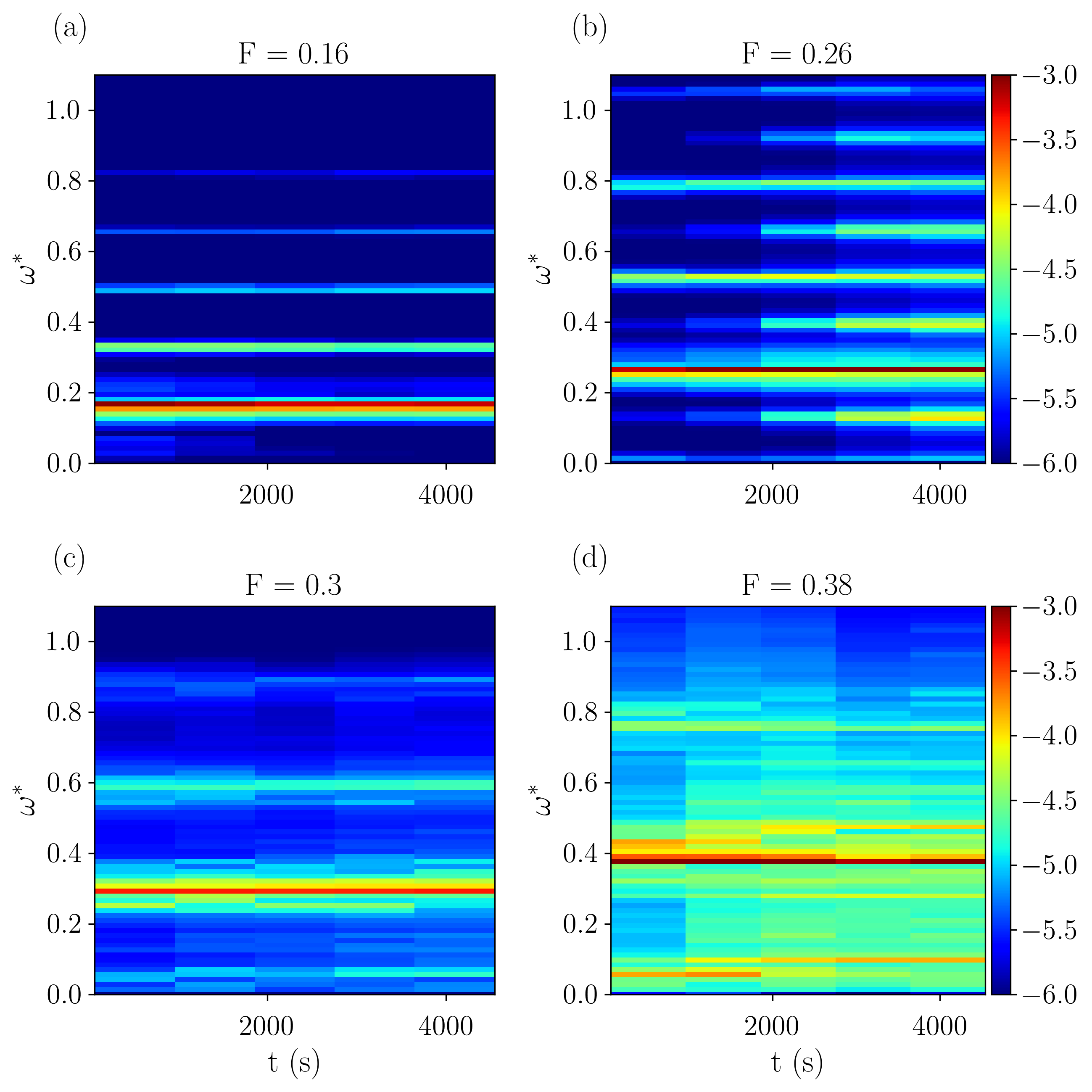}
	\caption{Spectrogram (time-frequency analysis): (a) EXP1, (b) EXP2, (c) EXP3, (d) EXP4. $t=0$ corresponds to approximately 20 minutes after starting the forcing.}
	\label{fig_spectrogram}
\end{figure}

For EXP2, with a higher forcing frequency $F=0.26$, the spectrum remains made of peaks but figure \ref{fig_PSD}(b) shows a richer energy spectrum, in which new energy peaks appear at $mF$ with $m=1/2, 1, 3/2, 2$ and so on. The spectrum has intermediate peaks in addition to the forcing harmonics already observed in EXP1. To study the time evolution of the energy spectra, the spectrogram is computed with consecutive Fourier transforms on five segments of the entire record (see figure \ref{fig_spectrogram}). We can notice that both in case (a) and (b), the forcing frequency and its harmonics do not evolve in time, conserving the same amount of energy over the entire duration of the experiment. For case (b), the frequencies $F/2$, $3F/2$...  show a time-dependent energy increase instead. {This growth over time of the half-harmonics suggests they are generated by energy transfer from nonlinear wave interactions. We speculate that they are due to a parametric subharmonic instability, a subset of the triadic resonance where the two daughter waves have very close frequencies $\omega_1 \approx \omega_2 \approx F/2$. Parametric subharmonic instability is often observed in the ocean in regions where the wave frequency is close to the local inertia frequency, and therefore, the forcing is particularly effective \cite{mackinnon2013parametric}. In our experiment, it is likely that the forcing frequency in EXP2 excites an internal resonance related to the pentagonal modes. \cite{touze2012transition} have observed a similar phenomenon in their experimental work on wave turbulence in vibrating plates, identifying a transition from a linear regime to a regime--named `quasi-periodic'--where half forcing frequencies arise before the final transition to full turbulence.} Waves at $F/2$ have the maximum growth rate, according to \cite{staquet2002internal} and provide a way for energy transfer towards small scales without a turbulent cascade process.

In the last two cases, figure \ref{fig_PSD}(c) for $F=0.30$ and (d) for $F=0.38$, the spectra show distinct behaviors below and above the $F$. 
First, the spectra become continuous even if some peaks, such as harmonics of the forcing peak, remain visible. In both cases, the spectrum is relatively flat for $\omega^\star<1$. In EXP3, most energy is contained in the poloidal component, which includes the waves. In contrast, almost no energy is visible in the toroidal part (except at a very low frequency, corresponding to the large-scale mode described in the previous part). The spectrum slightly extends at frequencies above $N$; however, the poloidal spectrum decays very fast (faster than $\omega^{-2}$) while the toroidal spectrum is at the noise level. By contrast, in EXP4, the toroidal spectrum remains weaker than the poloidal but is clearly above noise. Furthermore, both poloidal and toroidal spectra expand above $N$, and their decay is comparable to $\omega^{-2}$. They become closer to each other at $\omega>N$ than for $\omega<N$.

As discussed by \cite{Asaro2000}, $N$ is the parameter separating waves from turbulence. According to this model, for frequencies $\omega<N$, the spectrum is due to internal gravity waves and follows the $\omega^{-2}$ power-law typical of the GM spectrum (when the forcing frequency is much smaller than $N$). Above $N$, internal gravity waves cannot freely propagate ($N$ is the upper bound of their dispersion relation), and the motions are instead due to turbulence. 
The typical case in the open ocean is when the wave spectrum is much more energetic than the turbulent spectrum, resulting in a drop in the energy spectrum at frequencies corresponding to $N$. If the system is forced with more energy, the turbulent spectrum can match the wave energy with the consequent disappearance of the drop in the spectra. 

In our case, the extension of the $-2$ power-law decay for $\omega^*>1$, occurring for EXP4, seems to indicate that turbulence at high frequencies has a higher energy level when the forcing frequency is increased. The observed turbulence evolves toward strongly nonlinear turbulence.

\subsection{Space-frequency analysis}
\begin{figure}[htb]
	\centering
	\includegraphics[width=\linewidth]{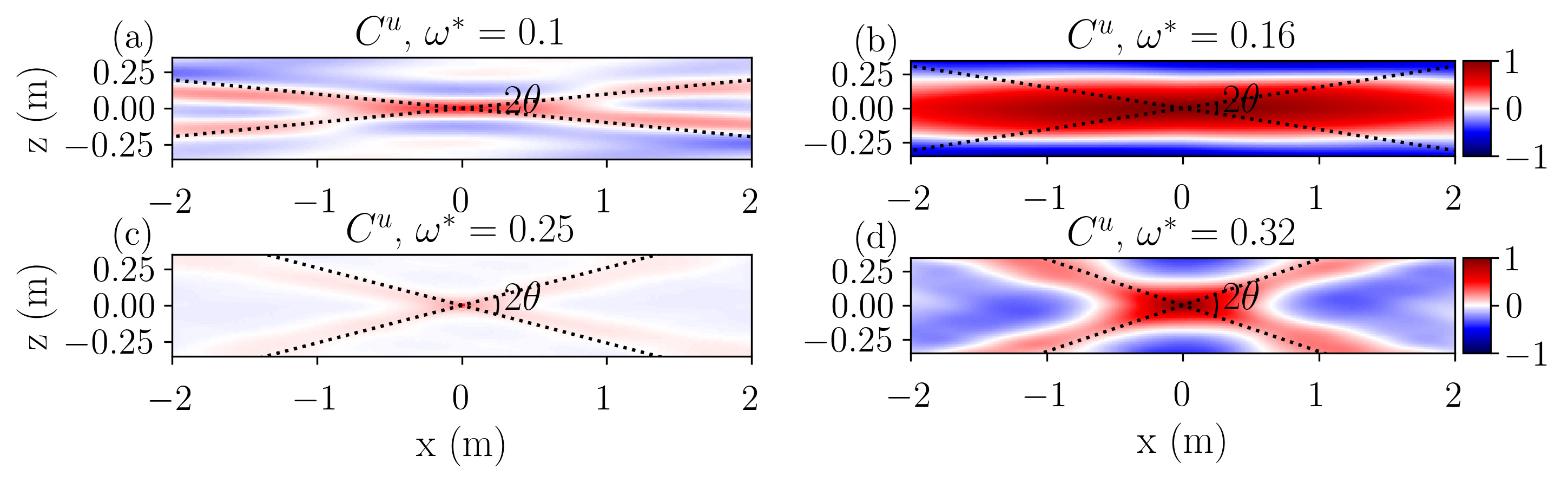}
	\caption{Autocorrelation $C^{u}(x,z,\omega)$ for EXP1 at four frequencies (see fig. \ref{fig_PSD}(a)): off-peak $\omega^* = 0.1< F$ (a),  on-peak $\omega^* =F$ (b), off-peak $F<\omega^*=0.25<1$ (c), and on-peak $\omega^*=2F$ (d). The black dashed line marks the dispersion relation for the frequency at which the autocorrelation is calculated. }
	\label{fig_exp1_autocorel}
\end{figure}
The spatial correlation $C^u$ for EXP1, calculated at four different frequencies, is shown in figure \ref{fig_exp1_autocorel}. The two panels on the left represent the plots at off-peak frequencies, below the forcing at $\omega^*=0.1$ in (a) and between the forcing and $N$ at $\omega^*=0.25$ in (c). The gravity wave dispersion relationship is well visible in both cases, with a diminished strength at higher frequencies. 
The correlation at the forcing, in figure \ref{fig_exp1_autocorel} (b), reveals a pattern similar to what was observed earlier for EXP0: a large-scale chessboard typical of a dominant wavelength wave.
Interestingly, for all the other frequencies, the on-peak correlation plots reveal a pattern following the internal wave dispersion relation with a higher magnitude than the off-peak patterns. Also, the cross arm's width is larger for the on-peak frequency, relating it to larger-scale waves. 
Other than at $F$, no chessboard patterns could be found for EXP1. It suggests that these peaks are not simply harmonics or the spatial mode at $F$ but have a complex spatial structure.\\

\begin{figure}[htb]
	\centering
	\includegraphics[width=\linewidth]{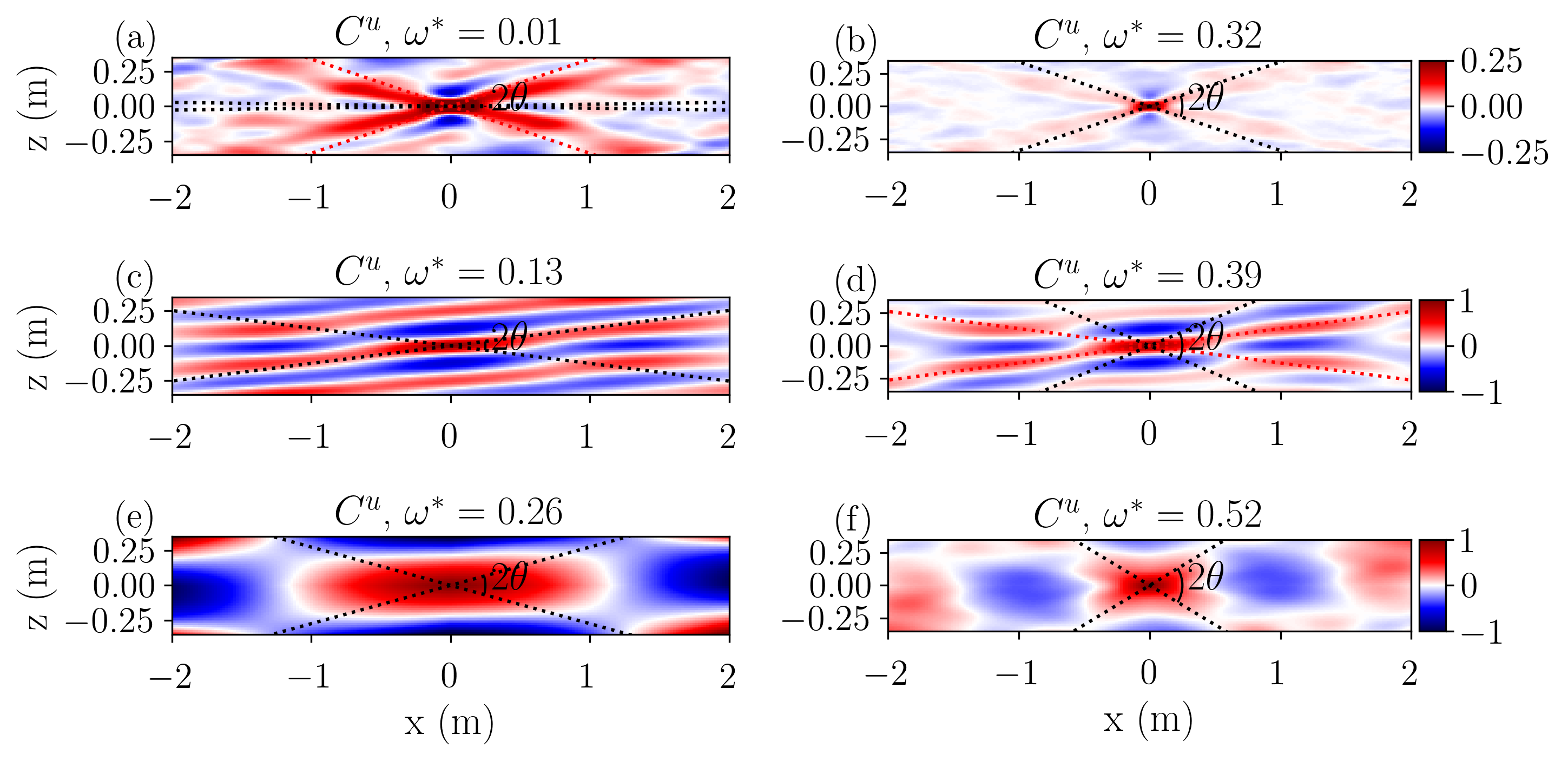}
	\caption{Autocorrelation $C^{u}(x,z,\omega)$ for EXP2 at six frequencies (see fig. \ref{fig_PSD}(b)): off-peak $\omega_0^* = 0.01< F$ (a) the red dotted line marks the dispersion relation for $0.01-F/2$, and off-peak $F<\omega_0^* = 0.32<1$ (b); on-peak at $\omega^*=F/2$ (c), and $\omega^*=3/2F$ (d); on-peak $\omega^*=F$ (d) the red dotted line marks the dispersion relation for $\omega^*=F/2$ (bound waves), and $\omega^*=2F$ (e). The black dashed line marks the dispersion relation for the frequency at which the autocorrelation is calculated.}
	\label{fig_exp2_autocorel}
\end{figure}

EXP2 has much richer dynamics, as shown in figure \ref{fig_exp2_autocorel}. Six different frequencies have been selected to represent the most significant cases. Off-peak frequencies are plotted in the top two panels, (a) and (b); on-peak at $\omega^*=F/2$ (c), and $\omega^*=3/2F$ (d) in the middle panels, while in the bottom panels are on-peak at $\omega^*=F$ (e), and $\omega^*=2F$ (f).
The off-peak has a cross again, but for the very low frequency (figure \ref{fig_exp2_autocorel}(a)), it corresponds to the dispersion relation for $0.01-F/2$ instead of the one for $\omega^*=0.01$. The phenomenon of bound waves has already been discussed in section \ref{sec:domain}, but in this case, the bound waves are observed for low frequencies rather than high ones.

At half the forcing and multiples, the correlation shows an unprecedented pattern. At $F/2$, in figure \ref{fig_exp2_autocorel}(c), the characteristic features are oblique lines with inclination corresponding to the dispersion relation. At $3/2F$, in figure \ref{fig_exp2_autocorel}(d), similar oblique lines can be seen, with the same inclination, i.e. at the dispersion relation for $F/2$ (highlighted by the red dotted line). This pattern is different from all the others observed in the experiments. The fact that only one arm of the cross is visible might indicate that internal waves with a preferential direction of propagation dominate at these frequencies.

At the forcing and its higher harmonics, shown in figure \ref{fig_exp2_autocorel}(e) and (f), we find the chessboard patterns again. Note that $k_x$ at $F$ is approximately half the value of $k_x$ for $2F$, hinting at a triadic resonance of various waves of similar frequencies close to $F$. In the red part at the middle of  \ref{fig_exp2_autocorel}(f), a short-armed cross can also be recognised when looked at carefully. This suggests a superposition of waves with different wavenumbers, although a dominant mode mostly determines the shape of the correlation.\\

\begin{figure}[htb]
	\centering
	\includegraphics[width=\linewidth]{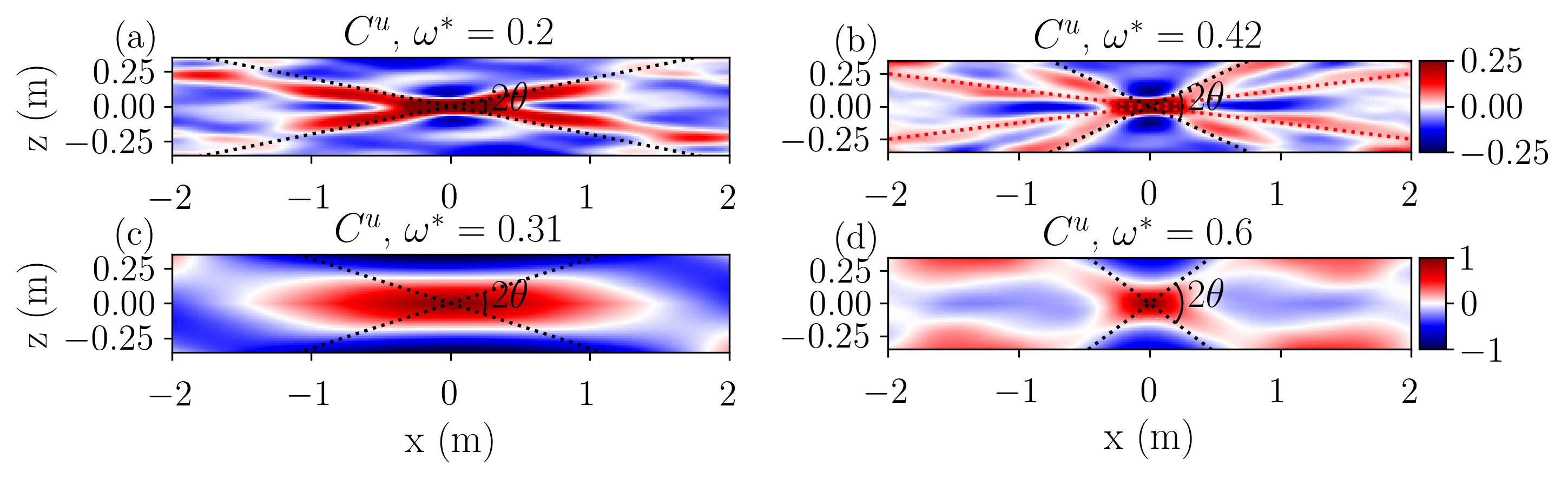}
	\caption{Autocorrelation $C^{u}(x,z,\omega)$ for EXP3 at four frequencies: off-peak $\omega^* = 0.2< F$ (a), and off-peak $F<\omega^*=0.42<1$ the red dotted line marks the dispersion relation for $\omega^*-F$ (b);  on-peak $F$ (c), and on-peak $\omega^*=2F$ (d). The black dashed line marks the dispersion relation for the frequency at which the autocorrelation is calculated.}
	\label{fig_exp3_autocorel}
\end{figure}

Despite the differences in frequency spectra, the spatial structures in EXP1 and EXP3 share some common features. At off-peak frequencies, see figures \ref{fig_exp3_autocorel} (a) for $\omega^*<F$, and (b) for  $F<\omega^*<1$, waves are making the continuum of the spectrum. In some cases, as at $\omega^*=0.42$, bound waves also play a role. At the forcing, the autocorrelation assumes a shape with a chessboard pattern that was already observed in the other experiments. However, as in EXP1, the higher harmonics, which are prominent peaks in the frequency spectra, are made by a random superposition of waves rather than a singular dominant wavelength. Both free and bound waves play a role in the dynamics of EXP3 for frequencies $F<\omega^*<1$, highlighting more complex dynamics. 
\begin{figure}[htb]
	\centering
	\includegraphics[width=\linewidth]{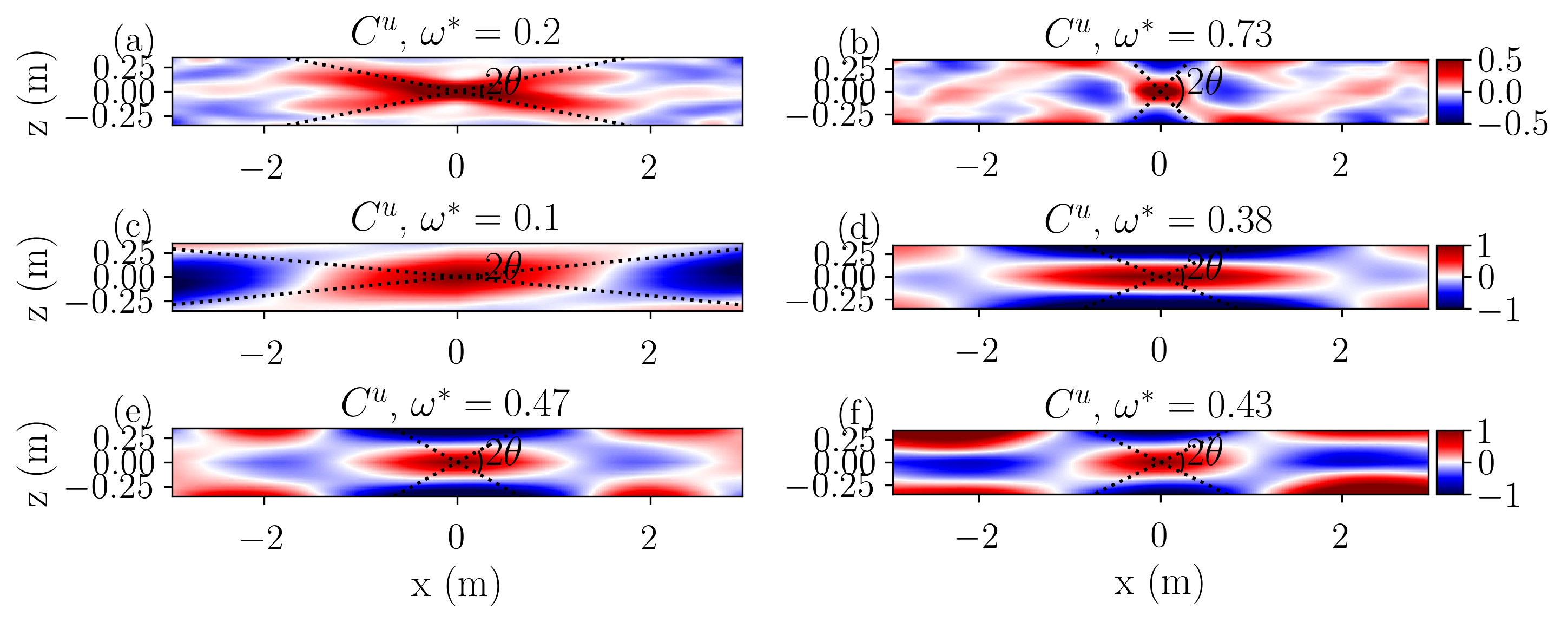}
	\caption{Autocorrelation $C^{u}(x,z,\omega)$ for EXP4 at six frequencies: off-peak $\omega^* = 0.2< F$ (a), and off-peak $F<\omega^* = 0.73<2F$ (b); on-peak at $\omega^*=0.1$ (c), on-peak $\omega^*=F$ (d); on-peak $\omega^*=0.47$ (e), on-peak $\omega^*=0.43$ (f). The black dashed line marks the dispersion relation for the frequency at which the autocorrelation is calculated.}
	\label{fig_exp4_autocorel}
\end{figure}

Finally, in EXP4 we can mainly identify two types of patterns: off-peak with visible dispersion relation both below and above $F$ (visible in figures \ref{fig_exp4_autocorel}(a) and (b)), and on-peak chess-board patterns (\ref{fig_exp4_autocorel}(c), (d), (e), and (f)). 
No signature of bound waves could be found in this experiment. The arm width of the cross in \ref{fig_exp4_autocorel}(a) is larger than the one in \ref{fig_exp4_autocorel}(b), indicating that the higher frequency waves are smaller-scale, similarly to what was observed for EXP0. The combined observation of a $\omega^{-2}$ slope extending above $N$ in the frequency spectra and a decreasing of the wave scales are an indicator of a turbulent cascade characterising this regime.

From comparing the spatial structures of the different experiments, it emerges that the continuum of the spectra (shown in the off-peaks correlation plots) at frequencies lower than the forcing $F$ is due to a superposition of internal gravity waves with different wavenumbers. In EXP2 and EXP3, some frequencies reveal bound waves, in addition to the free waves. 
At the most energetic frequency peaks, we have two distinct cases: there is a superposition of waves for the forcing higher harmonics, while there seems to be a single dominating spatial structure in the case of peaks corresponding to triadic resonances. The appearance of such dominating structure might be related to the triads coinciding with the modes of the pentagon.


\subsection{Nonlinear coupling analysis}
\begin{figure}[htb]
	\centering
	\includegraphics[width=0.9\linewidth]{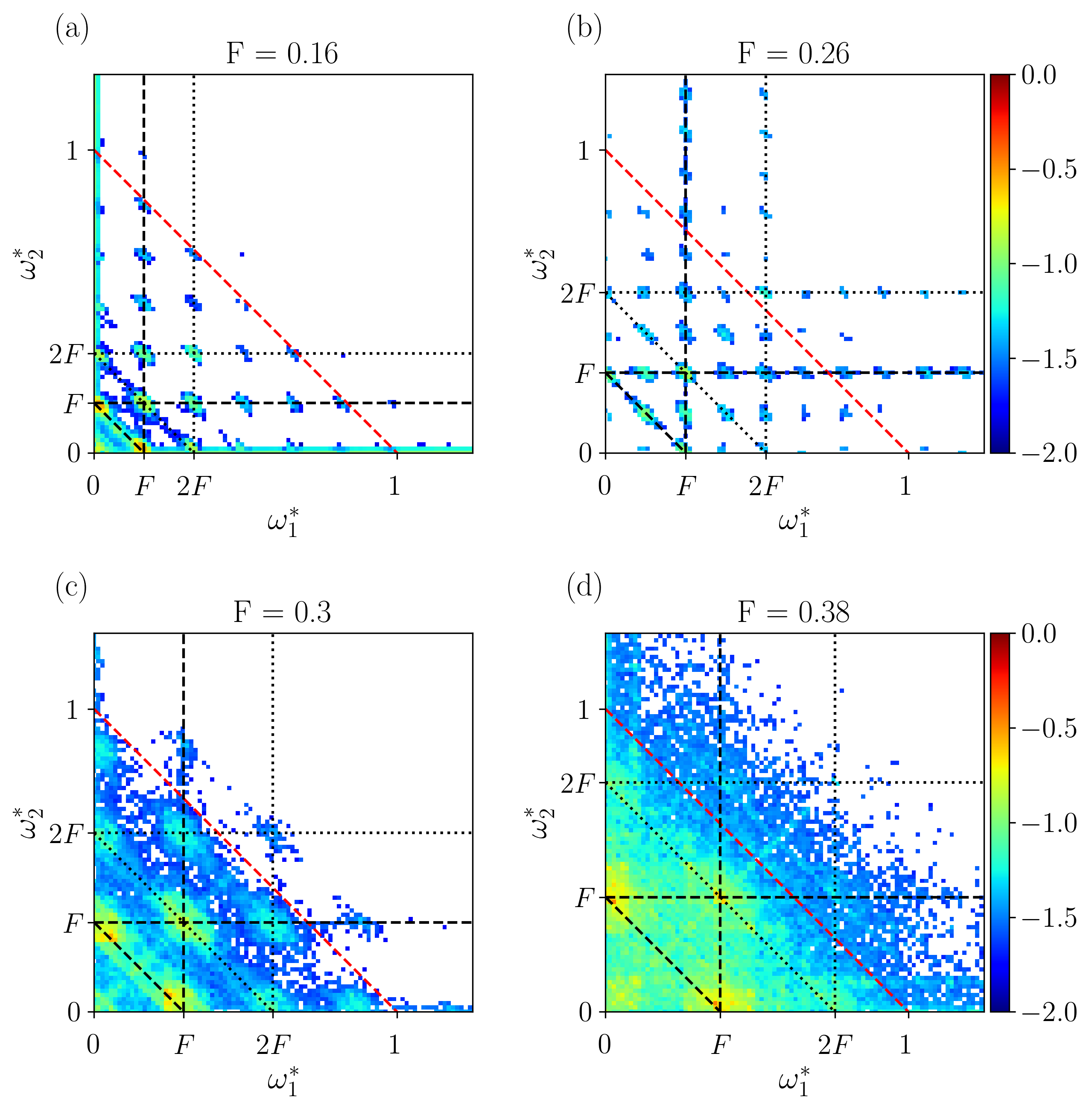}
	\caption{Bicoherence $B^u(\omega_1,\omega_2)$ for different experiments (in logarithmic scale): (a) EXP1, (b) EXP2, (c) EXP3, (d) EXP4. The dashed and dotted lines show the triads related to the forcing frequency $F$ and the first harmonic $2F$, respectively. The red dashed line marks $N$. As for fig. \ref{fig_control_bicoherence}, white corresponds to not statistically relevant values.}
	\label{fig_bicoherences}
\end{figure}

We can now look at the bicoherence (see the contour plots in figure \ref{fig_bicoherences}) to further investigate the wave interactions arising in the different experiments. For EXP1, figure \ref{fig_bicoherences}(a) shows only peaks at the frequencies $F$, $2F$ and $3F$ (the dashed and dotted lines are added to the plot to help identify the triads). This confirms that the waves are phase-coupled, and indeed triadic resonances cause the higher harmonics in the flow. 

Similarly, for EXP2 (figure \ref{fig_bicoherences}(b)), we have peaks at the integer multiples of the forcing frequency, and in addition, also peaks at half-integer multiples of the forcing. This indicates that secondary waves act as primary waves for higher-order triadic resonances, resulting in a cascade of triadic interactions. Such interactions transfer energy from the forced large-scale and large-amplitude waves towards internal waves with a smaller amplitude and smaller scale, where dissipation occurs. The bicoherence of EXP2 shows a discrete plot, consistent with the frequency spectra not displaying any continuous energy cascade. 
This is a characteristic feature of discrete wave turbulence \citep{katarastova2009}.

For EXP3, in figure \ref{fig_bicoherences} (c), the bicoherence is a continuous function. Some triadic resonances are still visible, particularly for the forcing and its higher harmonics. The peaks are very broad due to nonlinear spectral widening, and the coherence is spread out. Furthermore, at frequencies below $N$, a continuum fills in the bicoherence between peaks. For frequencies above $N$, the coherence level is not significant. This supports the fact that the motion is made of internal wave turbulence in this regime.

For the highest forcing frequency experiment EXP4, figure \ref{fig_bicoherences} (d), the bicoherence is a smooth function without peaks. In this case, the continuum is visible even for frequencies higher than $N$ (marked by the red dashed line in figure \ref{fig_bicoherences}), even though it is less important than in EXP0 (see figure \ref{fig_control_bicoherence}). The dynamics at frequencies above $N$ is significant and is not made of wave motions.

From the bicoherence analysis, we can conclude that, by increasing the frequency at which internal waves are forced, a transition from a discrete wave-interaction regime to a kinetic-like weak turbulence regime occurs. A similar scenario has been observed by \cite{Monsalve2020} for inertial wave turbulence. Still, the difference is that in our case, the transition is not due to an increase in $Re$, which remains approximately constant, but by the rise in $F_h$ and $Re_b$.

\section{Discussion of the dynamical regimes}
\label{Dyn}
In the previous sections, we have identified a first regime transition occurring at $\Re_b \approx 1$. The discrete wave turbulence regime appears for $Re_b<1$, where nonlinearity is very weak, and interactions only occur at a few discrete frequencies, i.e. at the frequency of the forced waves and a few harmonics or subharmonics. 
At $1\lesssim Re_b \lesssim 3.5$, non-linear coupling among waves is stronger (but still weakly nonlinear), leading to a coupling of all waves as expected in the weak turbulence phenomenology \citep{nazarenko2011wave}. The spectrum is continuous. However, the forcing frequency can not be lowered enough to have a large-scale separation $F\ll 1$ so that to be in the conditions of the observation of the Garrett \& Munk spectrum. Nevertheless, an evolution toward a decay of the spectrum as $\omega^{-2}$ is observed near the transition to strongly non-linear turbulence. In this regime, nonlinearity is strong enough so that to develop a weak nonlinear wave cascade but not strong enough to develop a strongly nonlinear cascade of vortices (notably in horizontal scales). Somehow it resembles to some extend the `viscosity-affected stratified flow regime' described in \citep{brethouwer2007} although $Re_b\sim 1$ rather than $Re_b\ll 1$ (allowing the wave cascade). Finally, the forcing (and possibly a few harmonics) peaks remain visible over the continuous spectrum, as is often the case in experiments (even in strongly nonlinear turbulence). 

For $Re_b\gtrsim 3.5$, the decay of the energy spectra extends beyond $N$ and has a slope comparable with the $\omega^{-2}$ GM spectra. The energy contained at frequencies above $N$ comes partly from high-frequency bound waves and vertical vorticity. Part of the poloidal energy is most likely due to horizontal vorticity, but further analyses must be performed to confirm it. Wave turbulence is present at low frequencies and large scales, while strongly nonlinear turbulence dominates at higher frequencies and smaller scales.

\cite{Asaro2000} reported on two dynamical regimes for mixing in the oceans: the wave regime, in which weak wave interactions cause the dynamics of the flow; and the turbulent regime, where waves play a minor role, and the evolution of the dynamics is driven by direct transfer from large scale controlled by instability and turbulence. Both regimes have been observed in different oceanic regions. The wave regime is typically observed in open oceans far from the coastline, while the turbulent regime is observed in regions of strong turbulence and mixing as the Knight Inlet near the Canadian coast. Comparing the oceanic spectra with the experimental ones, we found some similarities that indicate a similar transition between the two regimes. However, at the moderate value of $Re_b$ reached here, waves are still present in the dynamics in the strongly nonlinear regime. The flow is not quite in the strongly stratified turbulence regime expected by \citep{brethouwer2007} at $Re_b\gg1$.

We now consider the entire experimental dataset to identify the regime transitions better. The parameters of the additional experiments are listed in table \ref{table_exps_all}. $Re$, $Fr_h$, and $Re_b$, as defined in (\ref{eq:numbers}) are indicated for each experiment. The experiments have been classified by observing the decay of the frequency spectral (full and split in its poloidal and toroidal parts). Following the previous analysis, three regimes are distinguished: DWT discrete wave turbulence (frequency spectra made of discrete peaks), WT wave turbulence (continuous frequency spectra with a steep decay of the poloidal component and very weak energy in the toroidal spectrum), and ST stratified turbulence (a significant amount of energy in the toroidal part and decay of spectra comparable to $\omega^{-2}$). The Reynolds number takes values in a relatively narrow range $1.3-7.8 \times 10^4$ independently from the regime (see table \ref{table_exps_all}). Therefore, $Re$ is not a suitable indicator for the regimes developing in the flow. Our data consistently show that $Re_b$ is a better non-dimensional number to distinguish among the different regimes, confirming the relevance of $Re_b$ for our experiments on stratified turbulence.
\begin{figure}[htb]
	\centering
	\includegraphics[width=0.7\linewidth]{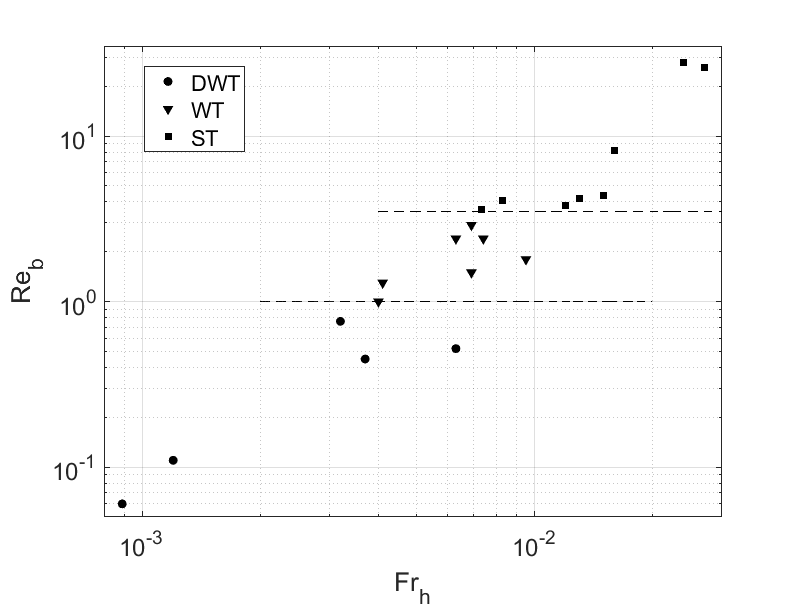}
	\caption{Phase diagram of our experiments as a function of the Froude number $Fr_h$ and the buoyancy Reynolds number $Re_b$. Round markers: discrete wave turbulence (DWT). Triangles: weak wave turbulence (WT). Squares: stratified turbulence (ST). Dashed lines are eye guides for the transitions. See table \ref{table_exps_all} for the detailed parameters.}
	\label{fig_Fr-Re}
\end{figure}

\section{Conclusions}
\label{sec:summary}

We report an experimental investigation of the dynamical regimes arising when large-scale internal waves force the flow in a stably stratified fluid.
We first investigated the consequences of changing the shape and size of the domain. The waves are forced with the same parameters as in \cite{savaro2020generation}, and the results obtained with the two set-ups are compared. In the square domain, finite-system size effects are strong and manifest in energy spectra characterised by discrete Fourier modes corresponding to the eigenmodes of the box. As expected, increasing the domain size reduces discreteness effects (as expected in the weak turbulence phenomenology \citep{nazarenko2011wave,falcon2022experiments}.

Three dynamical regimes have been identified when increasing the buoyancy Reynolds number: discrete wave turbulence at $Re_b < 1$, weak wave turbulence in a narrow range $1\lesssim Re_b \lesssim 3.5$ and at large $Re_b$ a strongly nonlinear regime that possibly evolves to strongly stratified turbulence \citep{Billant2001,brethouwer2007}.

The transition between the two strongest regimes in terms of wavelength is unclear and remains to be explored. For frequencies below $N$, a transition between both regimes may also occur in length scales when the linear timescale becomes comparable to the nonlinear timescale as a critical balance as observed in magnetohydrodynamics \cite{nazarenko2011wave}. A hint of this transition in length scales was reported in \cite{savaro2020generation} in their figures 11 \& 12. They observe that the small-scale cutoff of the space-time wave spectrum increases when forcing amplitude increases. This suggests that the transition between the wave regime and the strong turbulence regime occurs at larger scales when increasing the forcing (at a given frequency). Further investigations of the small lengthscale properties of the flow are required to characterise this transition.

\begin{acknowledgments}
This project has received financial support from Fondation Simone et Cino Del Duca of the Institut de France, from the European Research Council (ERC) under the European Union's Horizon 2020 research and innovation program (grant agreement No 647018-WATU) and from the Simons Foundation through the Simons collaboration on wave turbulence. 
\end{acknowledgments}

\bibliography{biblio3}

\end{document}